\documentclass[12pt, a4paper]{article}
\usepackage{graphicx}
\usepackage{amssymb}
\usepackage{amsmath}
\usepackage{bm}
\usepackage{color}
\usepackage{theorem}
\usepackage{subcaption}
\usepackage{listings}
\usepackage{colortbl}
\usepackage{tabularx}
\usepackage{longtable}
\usepackage[utf8]{inputenc}

\usepackage[sort&compress,numbers, merge]{natbib}

\definecolor{Orange}{cmyk}{0,0.61,0.87,0}
\definecolor{JungleGreen}{cmyk}{0.99,0,0.52,0}
\definecolor{OliveGreen}{cmyk}{0.64,0,0.95,0.40}
\definecolor{Brown}{cmyk}{0,0.81,1,0.60}
\definecolor{RoyalBlue}{cmyk}{0.71,0.53,0,0.12}
\definecolor{Gray}{cmyk}{0,0,0,0.40}
\definecolor{LightPink}{cmyk}{0.0,0.25,0,0}
\definecolor{LLightPink}{cmyk}{0.0,0.10,0,0}
\definecolor{LightBlue}{cmyk}{0.25,0,0,0}
\definecolor{LightGray}{cmyk}{0,0,0,0.2}

\newcommand{\1}{\mbox{1}\hspace{-0.25em}\mbox{l}}

\allowdisplaybreaks[1]

\usepackage[colorlinks=true, linkcolor=OliveGreen, citecolor=RoyalBlue,
urlcolor=RoyalBlue]{hyperref}

\renewcommand{\thefootnote}{\fnsymbol{footnote}}

\begin{document}

\begin{titlepage}

  \begin{flushright}
    {\tt
      UT-18-28\\ IPMU18-0191\\ KUNS-2740
    }
\end{flushright}

\vskip 1.35cm
\begin{center}

{\large
{\bf
Minimal Gauged $\text{U}(1)_{L_\alpha - L_\beta}$ Models Driven into a Corner
}
}

\vskip 1.5cm

Kento Asai$^{a}$\footnote{
\href{mailto:asai@hep-th.phys.s.u-tokyo.ac.jp}{\tt
 asai@hep-th.phys.s.u-tokyo.ac.jp}}, 
Koichi Hamaguchi$^{a,b}$\footnote{
\href{mailto:hama@hep-th.phys.s.u-tokyo.ac.jp}{\tt
 hama@hep-th.phys.s.u-tokyo.ac.jp}},
Natsumi Nagata$^a$\footnote{
\href{mailto:natsumi@hep-th.phys.s.u-tokyo.ac.jp}{\tt
 natsumi@hep-th.phys.s.u-tokyo.ac.jp}},\\[3pt]
Shih-Yen Tseng$^{a}$\footnote{
\href{mailto:shihyen@hep-th.phys.s.u-tokyo.ac.jp}{\tt
 shihyen@hep-th.phys.s.u-tokyo.ac.jp}},
and 
Koji Tsumura$^{c}$\footnote{
\href{mailto:ko2@gauge.scphys.kyoto-u.ac.jp}{\tt
 ko2@gauge.scphys.kyoto-u.ac.jp}}

\vskip 0.8cm

{\it $^a$Department of Physics, University of Tokyo, Bunkyo-ku, Tokyo
 113--0033, Japan} \\[2pt]
{\it ${}^b$Kavli IPMU (WPI), University of Tokyo, Kashiwa, Chiba 277--8583, Japan} \\[2pt]
{\it ${}^c$Department of Physics, Kyoto University, Kyoto 606--8502, Japan} 

\date{\today}

\vskip 1.2cm

\begin{abstract}

 It is well known that the differences between the lepton numbers can be
 gauged with the Standard Model matter content. Such extended gauge
 theories, dubbed as the gauged $\text{U}(1)_{L_\alpha - L_\beta}$ models,
 have been widely discussed so far as potential candidates for physics
 beyond the Standard Model. In this work, we study the
 minimal versions of these gauge theories, where three right-handed
 neutrinos as well as a single $\text{U}(1)_{L_\alpha - L_\beta}$ symmetry
 breaking Higgs field---an SU(2)$_L$ singlet or doublet---are
 introduced. In these minimal models, the neutrino mass terms are
 constrained by the gauge symmetry, which result in the two-zero
 texture or two-zero minor structure of neutrino mass matrices.
 Such restrictive forms of neutrino mass matrices lead to
 non-trivial predictions for the neutrino oscillation parameters as well
 as the size of the mass eigenvalues. We find that due to this restriction
 the minimal gauged $\text{U}(1)_{L_\alpha - L_\beta}$ models
 are either incompatible with the observed values of the neutrino parameters
 or in strong tension with the Planck 2018 limit on the
 sum of the neutrino masses. Only the U(1)$_{L_\mu-L_\tau}$ model with
 an SU(2)$_L$ singlet U(1)$_{L_\mu-L_\tau}$-breaking field barely evades
 the limit, which can be tested in the future neutrino experiments.

\end{abstract}

\end{center}
\end{titlepage}

\renewcommand{\thefootnote}{\arabic{footnote}}
\setcounter{footnote}{0}

\section{Introduction}
\label{sec:intro}

The Standard Model (SM) of particle physics proved extremely successful
in describing most of the phenomena below the TeV scale. This is based
on the $\text{SU}(3)_C \otimes \text{SU}(2)_L \otimes \text{U}(1)_Y$
gauge theory with three generations of quarks and leptons as well as one
$\text{SU}(2)_L$ doublet Higgs field. Although this gauge structure seems
necessary and sufficient to explain most of the experimental data so
far, the SM potentially allows an extension of the gauge sector by
gauging one of the accidental U(1) symmetries in the SM
\cite{Foot:1990mn, He:1990pn, He:1991qd, Foot:1994vd}. Among the
possibilities of such U(1) symmetries, the differences in the lepton
numbers are frequently considered in various contexts. We denote these
symmetries by U(1)$_{L_\alpha-L_\beta}$, where
$L_\alpha$ ($\alpha = e,\mu,\tau$) represent the lepton number for each
flavor. In particular, the U(1)$_{L_\mu-L_\tau}$ models are quite motivated since the
new gauge interaction mediated by the U(1)$_{L_\mu-L_\tau}$ gauge boson
may account for the muon $g-2$ anomaly \cite{Bennett:2006fi,
Jegerlehner:2009ry, Davier:2010nc, Hagiwara:2011af} while avoiding the
experimental constraints thanks to the absence of its couplings with
electron and quarks at tree level \cite{Baek:2001kca,
Ma:2001md}. In addition, possibilities of explaining
flavor anomalies with this gauge boson have also been discussed
\cite{Altmannshofer:2014cfa, Crivellin:2015mga}. The
U(1)$_{L_\mu-L_\tau}$ gauge boson is often utilized also in dark matter models
in order to realize the correct dark matter abundance
\cite{Kim:2015fpa, Baek:2015fea, Patra:2016shz, Biswas:2016yan,
Biswas:2016yjr, Biswas:2017ait, Foldenauer:2018zrz}. 
Other recent related studies on the gauged U(1)$_{L_\alpha -L_\beta}$ models are found in
Refs.~\cite{Heeck:2011wj, Harigaya:2013twa, delAguila:2014soa,
Fuyuto:2014cya, Araki:2015mya, 
Elahi:2015vzh, Fuyuto:2015gmk, Altmannshofer:2016oaq, Ibe:2016dir,
Kaneta:2016uyt, Araki:2017wyg, Hou:2017ozb, Chen:2017cic,
Elahi:2017ppe, Chen:2017gvf, Baek:2017sew, Gninenko:2018tlp,
Wise:2018rnb,
Nomura:2018yej, Nomura:2018vfz, Bauer:2018onh, Arcadi:2018tly, Kamada:2018zxi,
Liu:2018xsw, Bauer:2018egk, Nomura:2018cle, Banerjee:2018eaf, Banerjee:2018mnw,
Chun:2018ibr}.

Under the U(1)$_{L_\alpha-L_\beta}$ gauge symmetries, only the lepton sector
is transformed non-trivially. This motivates us to study if the
lepton sector of the gauged U(1)$_{L_\alpha-L_\beta}$ models is compatible
with the existing experimental results. In particular, the models should
account for the observed pattern of neutrino oscillations, which
constrains possible flavor structures of the lepton sector. For previous
studies on the neutrino sector of the gauged U(1)$_{L_\alpha-L_\beta}$
models, see Refs.~\cite{Branco:1988ex, Choubey:2004hn, Araki:2012ip,
Heeck:2014sna, Baek:2015mna, Crivellin:2015lwa, Plestid:2016esp, 
Lee:2017ekw, Asai:2017ryy, Dev:2017fdz}. 
To obtain a successful model, it is required to introduce right-handed
neutrinos. These right-handed neutrinos can have the Dirac mass terms
with left-handed neutrinos as well as the Majorana mass terms among
themselves, and if the size of the former is much smaller than that of the
latter, the Type-I seesaw mechanism \cite{Minkowski:1977sc,
Yanagida:1979as, GellMann:1980vs, Mohapatra:1979ia} generates small
masses for active neutrinos.  It however turns out that the introduction
of right-handed neutrinos is insufficient, as
the U(1)$_{L_\alpha-L_\beta}$ gauge symmetries forbid many of the Dirac and
Majorana mass terms, forcing the neutrino mass matrix to be
block-diagonal. Such a block-diagonal neutrino mass matrix is unable to
explain the neutrino oscillation data. We thus need to break these gauge
symmetries spontaneously by using vacuum expectation values (VEVs) of
additional scalar fields. For the VEV of a scalar field to affect the
neutrino mass structure through the renormalizable interactions, it
should be an SU(2)$_L$ singlet with hypercharge zero
or doublet with hypercharge $1/2$.\footnote{
An SU(2)$_L$ triplet scalar with hypercharge one, which can couple
to bilinear terms of the doublet leptons, may also be introduced. 
We however find that the resultant neutrino mass structure is 
more restrictive than those considered in this paper and thus 
unable to reproduce the observed pattern of neutrino mixing. } 
It is then found that if such a scalar field
has the U(1)$_{L_\alpha-L_\beta}$ charge $\pm 1$ and there are more than
three right-handed neutrinos, all of the three active neutrinos can
mix with each other. Therefore, the simplest and potentially viable
U(1)$_{L_\alpha-L_\beta}$ models consist of three right-handed neutrinos and a
single U(1)$_{L_\alpha-L_\beta}$-breaking scalar field besides the SM matter
fields. We refer to such models as the minimal gauged
U(1)$_{L_\alpha-L_\beta}$ models and focus on them in this paper.

In the minimal gauged U(1)$_{L_\alpha-L_\beta}$ models, there is still a
strong constraint on the neutrino mass structure, since a single
U(1)$_{L_\alpha-L_\beta}$-breaking scalar cannot give rise to all of the
components in the neutrino mass matrices \cite{Branco:1988ex,
Choubey:2004hn, Araki:2012ip, Heeck:2014sna, Baek:2015mna,
Crivellin:2015lwa, Plestid:2016esp, Lee:2017ekw, Asai:2017ryy,
Dev:2017fdz}. It is found that these models have a neutrino mass matrix
that has the form of the so-called two-zero texture \cite{Berger:2000zj,
Frampton:2002yf, Xing:2002ta, Kageyama:2002zw, Xing:2002ap} or two-zero
minor \cite{Lavoura:2004tu, Lashin:2007dm} structure. These structures
require the low-energy neutrino parameters (three mass eigenvalues,
three mixing angles, and three CP phases) to satisfy two conditional
expressions, which make four parameters among them dependent
on the rest of the parameters. Therefore, given the observed values of
the neutrino mixing angles and the mass-squared differences, we can
predict the Dirac and Majorana CP phases as well as the mass eigenvalue
of the lightest state by solving the conditional equations. However,
it should be noted that these equations may have no viable solution; in fact, it was shown in
Ref.~\cite{Asai:2017ryy} that among the minimal gauged
U(1)$_{L_\alpha-L_\beta}$ models with a singlet
U(1)$_{L_\alpha-L_\beta}$-breaking scalar boson, we could find a
solution only for the U(1)$_{L_\mu-L_\tau}$ model. 
In addition, with the neutrino oscillation parameters obtained from the global fit that was the latest then, 
it was found that for the U(1)$_{L_\mu-L_\tau}$ model the Dirac CP
phase was predicted in an experimentally favored range, while
the sum of the neutrino masses turned out to be rather large,
$0.12$--0.40~eV, and a part of this range had already been disfavored by
the Planck 2015 result, $\sum_{i} m_i < 0.23$~eV \cite{Ade:2015xua}. 

These results, however, need to be reconsidered given that several
new data have been published after the work. Among other things, there are two
important updates. First, the favored range of $\theta_{23}$, on which
the above predictions have strong dependence, changed from the previous
result in the up-to-date global fit given by \texttt{NuFIT v4.0}
\cite{nufit, Esteban:2018azc}. This is mainly because the value of
$\theta_{23}$ favored by the NOvA experiment shifted \cite{NOvA:2018gge}
from the previous value \cite{Adamson:2017gxd}, which is now in good
agreement with the T2K result \cite{Abe:2017vif, Abe:2018wpn}. Second,
the Planck collaboration reported new results for the measurements of
the cosmological parameters and gave a more stringent upper limit on the
sum of the neutrino masses: $\sum_{i} m_i < 0.12$~eV
\cite{Aghanim:2018eyx}. By comparing this with the aforementioned values
predicted in Ref.~\cite{Asai:2017ryy}, we see that there is a tension
between them. 

Motivated by this demand, in this paper, we study the structure of the
neutrino mass matrices in the minimal gauged U(1)$_{L_\alpha-L_\beta}$
models. We discuss not only the case with an SU(2)$_L$ singlet
U(1)$_{L_\alpha-L_\beta}$-breaking scalar boson as in
Ref.~\cite{Asai:2017ryy}, but also those with SU(2)$_L$ doublet ones. We
list all of the minimal gauged U(1)$_{L_\alpha-L_\beta}$ models and
systematically check their predictions against the latest experimental
results. It is found that all of the models except for the
U(1)$_{L_\mu-L_\tau}$ model with a singlet scalar have already been excluded. The
remaining one possibility is also in a rather strong tension with the
Planck 2018 limit on $\sum_{i} m_i$, and this case will soon be tested
in the future neutrino experiments.

This paper is organized as follows. In the subsequent section, we list
all of the possible minimal gauged U(1)$_{L_\alpha-L_\beta}$ models and show
their particle content, the assignment of quantum numbers, and the
Lagrangian terms relevant to the neutrino mass matrices. It turns out
that the SU(2)$_L$ doublet cases accommodate the charged lepton flavor
violation---we study the phenomenological consequences of such effect in
Sec.~\ref{sec:lfv}. In Sec.~\ref{sec:neumass}, we show the structure of
the neutrino mass matrix in each model, and derive the conditional
expressions imposed on the low-energy neutrino parameters. We then study the
prediction for the sum of the neutrino masses obtained in each model in
Sec.~\ref{sec:limit}, and compare them to the Planck limit. For the
U(1)$_{L_\mu-L_\tau}$ model with a singlet scalar, which is the only
case that has not been completely excluded yet, we also show other predictions
and discuss the testability of this model. Finally,
Sec.~\ref{sec:conclusion} is devoted to conclusion and discussion.

\section{Models}
\label{sec:models}

To formulate the minimal gauged U(1)$_{L_\alpha-L_\beta}$ models, we
first define the lepton flavors in our setup. Usually, the flavors of
charged leptons are defined with respect to their mass eigenstates. In
the U(1)$_{L_\alpha-L_\beta}$ gauge theories, however, the assignment of
the gauge charges itself distinguishes the lepton flavors, and thus it
is more convenient to define the lepton flavors in the gauge
eigenbasis. The difference between these two definitions becomes
manifest when the Dirac mass matrix for the charged leptons in the gauge
eigenbasis is different from $\text{diag}(m_e, m_\mu, m_\tau)$, where
$m_e$, $m_\mu$, and $m_\tau$ are the masses of electron, muon, and tau,
respectively---we will note the difference when we consider
such a situation in the following discussions.

In this work, we consider a U(1) gauge theory where the three
flavors of the charged leptons have the U(1) charges of 0, $+1$, and
$-1$. We refer to these charged leptons as $e$, $\mu$, and $\tau$,
respectively, without loss of generality. As this assignment is
equivalent to $L_\mu - L_\tau$ in the ordinary sense, we call this
symmetry the U(1)$_{L_\mu-L_\tau}$ gauge symmetry. 
In this formulation, the other U(1)$_{L_\alpha-L_\beta}$ gauge theories
in the mass eigenbasis are obtained when the charged lepton mass matrix
has a form of
\begin{equation}
 M_\ell = D_3 (g)
\begin{pmatrix}
 m_e & 0& 0\\
 0 & m_\mu & 0 \\
 0 & 0 & m_\tau 
\end{pmatrix}
D_3^T (g) ~,
\label{eq:mlgeneric}
\end{equation}
where $D_3(g)$ denotes a three-dimensional real
representation of the symmetry group $S_3$, with $g$ an element of the
group: $g\in S_3$. Through this equation there is a one-to-one correspondence
between an element of $g \in S_3$ and the diagonal components of $M_\ell
= \text{diag}(m_{\ell_1}, m_{\ell_2}, m_{\ell_3})$; we thus denote the
element by $g_{{\ell_1}{\ell_2}{\ell_3}}$. The U(1)$_{L_\alpha-L_\beta}$
gauge theory is then obtained for $g_{{\ell_1}{\ell_2}{\ell_3}}$ that
transforms $\mu \to \alpha$, $\tau \to \beta$, and $e$ into the
remaining flavor. 

The left-handed neutrino $\nu_\ell$ has the same U(1)$_{L_\mu-L_\tau}$
gauge charge as that of the charged lepton counterpart, $\ell_{L, R}$. 
In addition, we introduce three right-handed neutrinos $N_e$, $N_\mu$,
and $N_\tau$, which have the U(1)$_{L_\mu - L_\tau}$ charges of $0$,
$+1$, and $-1$, respectively. All of the SM quarks are
not charged under the U(1)$_{L_\mu-L_\tau}$ gauge symmetry. With this
choice of quantum numbers, the theory is free from gauge anomalies
\cite{Foot:1990mn, He:1990pn, He:1991qd, Foot:1994vd}.\footnote{In fact,
with three right-handed neutrinos, there are more options for an extra
gauge symmetry than those discussed in Refs.~\cite{Foot:1990mn,
He:1990pn, He:1991qd, Foot:1994vd}. For a comprehensive discussion about
this, see Refs.~\cite{Liu:2011dh, Araki:2012ip, Kownacki:2016pmx, Tang:2017gkz}. }
We also exploit an SM(-like) Higgs field, \textit{i.e.} 
an SU(2)$_L$ doublet scalar with hypercharge $+1/2$
and the U(1)$_{L_\mu-L_\tau}$ charge zero; this
scalar field is responsible for giving masses to the SM fields.

As we discussed in the previous section, we further introduce one extra scalar
field to break the U(1)$_{L_\mu-L_\tau}$ gauge symmetry. 
There are only three possibilities for the quantum numbers of the
scalar field that can yield a neutrino mass matrix
with which all of the three active neutrinos mix with each other: 
\begin{itemize}
 \item[(i)] An SU(2)$_L$ singlet with hypercharge $Y=0$ and the
       U(1)$_{L_\mu-L_\tau}$ charge $+1$. 
 \item[(ii)] An SU(2)$_L$ doublet with hypercharge $Y=1/2$ and the
	     U(1)$_{L_\mu-L_\tau}$ charge $+1$.
 \item[(iii)] An SU(2)$_L$ doublet with hypercharge $Y=1/2$ and the
	     U(1)$_{L_\mu-L_\tau}$ charge $-1$.
\end{itemize}
For the case (i), one may also think of the U(1)$_{L_\mu-L_\tau}$ charge
$-1$ case. However, this case is just the complex conjugate of the case
(i) and thus these two are equivalent. Similarly, the choice of $Y =
-1/2$ in the cases of (ii) and (iii) is the complex conjugate of the
cases (iii) and (ii), respectively.

In what follows, we discuss each case separately, showing the Lagrangian
terms relevant to the neutrino mass structure.

\subsection{Singlet}

The interaction terms in the leptonic sector of the case (i) are given by
\begin{align}
 \Delta {\cal L} = 
& -y_e e_R^c L_e H^\dagger 
-y_\mu \mu_R^c L_\mu H^\dagger 
-y_\tau \tau_R^c L_\tau H^\dagger 
\nonumber \\[2pt]
&-\lambda_e N_e^c (L_e \cdot H)
-\lambda_\mu N_\mu^c (L_\mu \cdot H)
-\lambda_\tau N_\tau^c (L_\tau \cdot H) \nonumber \\[2pt]
&-\frac{1}{2}M_{ee} N_e^c N_e^c 
- M_{\mu \tau} N_\mu^c N_\tau^c 
- \lambda_{e\mu} \sigma N_e^c N_\mu^c
- \lambda_{e\tau} \sigma^* N_e^c N_\tau^c +\text{h.c.} ~,
\label{eq:lag}
\end{align}
where $H$ and $\sigma$ denote the SM Higgs and the
U(1)$_{L_\mu-L_\tau}$-breaking singlet scalar, respectively,
and $L_\alpha$ are the left-handed lepton doublets. 
The dots indicate the contraction of the SU(2)$_L$ indices.
After the Higgs field $H$ and the singlet scalar $\sigma$ acquire VEVs
$\langle H \rangle = v/\sqrt{2}$ and $\langle \sigma
\rangle$,\footnote{We can always take these VEVs to be real by
using the gauge transformations. }
respectively, these interaction terms lead to neutrino mass terms
\begin{align}
 {\cal L}_{\rm mass}^{(N)} &= -(\nu_e, \nu_\mu, \nu_\tau) {M}_D 
\begin{pmatrix}
N_e^c\\ N_\mu^c\\ N_\tau^c 
\end{pmatrix}
- \frac{1}{2}(N_e^c, N_\mu^c, N_\tau^c) {M}_R 
\begin{pmatrix}
 N_e^c \\ N_\mu^c \\ N_\tau^c 
\end{pmatrix}
+\text{h.c.} ~,
\end{align}
with
\begin{equation}
 {M}_D = \frac{v}{\sqrt{2}}
\begin{pmatrix}
 \lambda_e & 0& 0\\
 0 & \lambda_\mu & 0 \\
 0 & 0 & \lambda_\tau 
\end{pmatrix}
~,\qquad
{M}_R =
\begin{pmatrix}
 M_{ee} & \lambda_{e\mu} \langle \sigma \rangle & \lambda_{e\tau} 
\langle \sigma \rangle \\
 \lambda_{e\mu} \langle \sigma \rangle & 0 & M_{\mu\tau} \\
\lambda_{e\tau} \langle \sigma \rangle & M_{\mu\tau} & 0
\end{pmatrix}
~,
\label{eq:mdrs}
\end{equation}
and the charged lepton mass terms
\begin{align}
 {\cal L}_{\rm mass}^{(L)} &= -(e_L, \mu_L, \tau_L) {M}_\ell
\begin{pmatrix}
e_R^c\\ \mu_R^c\\ \tau_R^c 
\end{pmatrix}
+\text{h.c.} ~,
\end{align}
with
\begin{equation}
 M_\ell =  \frac{v}{\sqrt{2}}
\begin{pmatrix}
 y_e & 0& 0\\
 0 & y_\mu & 0 \\
 0 & 0 & y_\tau 
\end{pmatrix}
~.
\label{eq:mls}
\end{equation}
It is found that both the neutrino and charged-lepton Dirac mass
matrices are diagonal---they are assured by the U(1)$_{L_\mu - L_\tau}$
gauge symmetry. All of the components of these Dirac mass matrices are
taken to be real and positive without loss of generality. In this basis,
the matrix $M_\ell$ in Eq.~\eqref{eq:mls} is in general has a
form~\eqref{eq:mlgeneric}.

In this model, the U(1)$_{L_\mu-L_\tau}$ breaking scale is set by the
VEV of $\sigma$. Since $\sigma$ is singlet under the SM gauge group,
this breaking scale can be much higher than the electroweak scale so
that the seesaw mechanism \cite{Minkowski:1977sc,
Yanagida:1979as, GellMann:1980vs, Mohapatra:1979ia} naturally explains
the smallness of the active neutrino masses. Another interesting
possibility for the U(1)$_{L_\mu-L_\tau}$ breaking scale is motivated by
the muon $g-2$ anomaly \cite{Bennett:2006fi, 
Jegerlehner:2009ry, Davier:2010nc, Hagiwara:2011af}. It is known that
the observed deviation in the anomalous magnetic dipole moment of the muon from
the SM prediction can be accounted for by the contribution of the
U(1)$_{L_\mu-L_\tau}$ gauge boson at one-loop level \cite{Baek:2001kca,
Ma:2001md} without conflicting with the existing experiments if the mass
of the gauge boson is $m_{Z^\prime} \sim 10-100$~MeV and the
U(1)$_{L_\mu-L_\tau}$ gauge coupling is $g_{Z^\prime} \sim (5-10)\times
10^{-4}$. The lower edge of the mass range is due to the limit imposed
by the Borexino experiment \cite{Bellini:2011rx}, which gives a bound on
the $\nu$-$e$ interactions induced at loop level in this
model. $m_{Z^\prime} \lesssim 10$~MeV is also disfavored in cosmology as
it contributes to the effective neutrino degrees of freedom
\cite{Kamada:2018zxi}. On the other hand, the large mass region
$m_{Z^\prime} \gtrsim 100$~MeV is constrained by the measurements of the
neutrino trident production process \cite{Geiregat:1990gz,
Mishra:1991bv, Altmannshofer:2014cfa, Altmannshofer:2014pba} and by the
BABAR experiment searching for $e\bar{e} \to \mu \bar{\mu} Z^\prime$,
$Z^\prime \to \mu \bar{\mu}$ \cite{TheBABAR:2016rlg}. Since the mass of
the U(1)$_{L_\mu-L_\tau}$ boson is given by $m_{Z^\prime} = \sqrt{2}
g_{Z^\prime} \langle \sigma \rangle$, the muon $g-2$ anomaly can be
explained for $\langle \sigma \rangle \sim 10-100$~GeV. In the following
discussion, however, we do not stick to this range but regard $\langle
\sigma \rangle$ as just a free parameter.

\subsection{Doublet with the U(1)$_{L_\mu-L_\tau}$ charge $+1$}
\label{sec:modeldpl}

The generic interaction Lagrangian in the lepton sector for the case
(ii) is given by
\begin{align}
 \Delta {\cal L} = 
& -y_e e_R^c L_e \Phi_2^\dagger 
-y_\mu \mu_R^c L_\mu \Phi_2^\dagger 
-y_\tau \tau_R^c L_\tau \Phi_2^\dagger 
 -y_{\mu e} e_R^c L_\mu \Phi_1^\dagger 
-y_{e\tau } \tau_R^c L_e \Phi_1^\dagger 
\nonumber \\[2pt]
&-\lambda_e N_e^c (L_e \cdot \Phi_2)
-\lambda_\mu N_\mu^c (L_\mu \cdot \Phi_2)
-\lambda_\tau N_\tau^c (L_\tau \cdot \Phi_2) \nonumber \\[2pt]
&-\lambda_{\tau e} N_e^c (L_\tau \cdot \Phi_1)
-\lambda_{e\mu } N_\mu^c (L_e \cdot \Phi_1)
-\frac{1}{2}M_{ee} N_e^c N_e^c 
- M_{\mu \tau} N_\mu^c N_\tau^c 
+\text{h.c.} ~,
\label{eq:lagdpl}
\end{align}
where $\Phi_1$ ($\Phi_2$) is an SU(2)$_L$ doublet scalar field with
hypercharge $1/2$ and the U(1)$_{L_\mu-L_\tau}$ charge $+1$ ($0$). We
denote the VEVs of these fields by\footnote{We can take both $v_1$ and
$v_2$ to be real and positive through gauge transformations without loss
of generality.}
\begin{equation}
 \langle \Phi_i \rangle = \frac{1}{\sqrt{2}} 
\begin{pmatrix}
 0 \\ v_i
\end{pmatrix}
~,
\end{equation}
for $i = 1,2$, and define $v \equiv \sqrt{v_1^2 + v_2^2}$. The Dirac and
Majorana neutrino mass matrices are then given by
\begin{equation}
 {M}_D = \frac{1}{\sqrt{2}}
\begin{pmatrix}
 \lambda_e v_2 &  \lambda_{e \mu } v_1 & 0\\
0 & \lambda_\mu v_2 & 0 \\
 \lambda_{\tau e} v_1 & 0 & \lambda_\tau v_2 
\end{pmatrix}
~,\qquad
{M}_R =
\begin{pmatrix}
 M_{ee} & 0 & 0 \\
0 & 0 & M_{\mu\tau} \\
0 & M_{\mu\tau} & 0
\end{pmatrix}
~,
\label{eq:mdrdpl}
\end{equation}
while for the charged lepton mass matrix we have
\begin{equation}
 M_\ell =  \frac{1}{\sqrt{2}}
\begin{pmatrix}
 y_e v_2 & 0& y_{e\tau} v_1\\
 y_{\mu e} v_1 & y_\mu v_2 & 0 \\
 0 & 0 & y_\tau v_2 
\end{pmatrix}
~.
\label{eq:mldpl}
\end{equation}
Notice that in this case $M_\ell$ has off-diagonal components. Their
effect on the charged lepton-flavor-violating processes is discussed in
Sec.~\ref{sec:lfv}.

Contrary to the previous case, the U(1)$_{L_\mu-L_\tau}$-symmetry
breaking scale, which is determined by the VEV $v_1$, is bounded
from above in the present case since $v_1$ should satisfy $v =
\sqrt{v_1^2 + v_2^2} \simeq 246$~GeV. Therefore, this setup predicts the
U(1)$_{L_\mu-L_\tau}$ gauge boson to have a mass below the electroweak
scale.

\subsection{Doublet with the U(1)$_{L_\mu-L_\tau}$ charge $-1$}

The relevant Lagrangian terms for the case (iii) are
\begin{align}
 \Delta {\cal L} = 
& -y_e e_R^c L_e \Phi_2^\dagger 
-y_\mu \mu_R^c L_\mu \Phi_2^\dagger 
-y_\tau \tau_R^c L_\tau \Phi_2^\dagger 
 -y_{\tau e} e_R^c L_\tau \Phi_1^\dagger 
-y_{e\mu } \mu_R^c L_e \Phi_1^\dagger 
\nonumber \\[2pt]
&-\lambda_e N_e^c (L_e \cdot \Phi_2)
-\lambda_\mu N_\mu^c (L_\mu \cdot \Phi_2)
-\lambda_\tau N_\tau^c (L_\tau \cdot \Phi_2) \nonumber \\[2pt]
&-\lambda_{\mu e} N_e^c (L_\mu \cdot \Phi_1)
-\lambda_{e\tau } N_\tau^c (L_e \cdot \Phi_1)
-\frac{1}{2}M_{ee} N_e^c N_e^c 
- M_{\mu \tau} N_\mu^c N_\tau^c 
+\text{h.c.} ~,
\label{eq:lagdmi}
\end{align}
where $\Phi_1$ ($\Phi_2$) is an SU(2)$_L$ doublet scalar field with
hypercharge $1/2$ and the U(1)$_{L_\mu-L_\tau}$ charge $-1$ ($0$). We
define the VEVs of these fields in the same way as above. The Dirac and
Majorana neutrino mass matrices are then given by
\begin{equation}
 {M}_D = \frac{1}{\sqrt{2}}
\begin{pmatrix}
 \lambda_e v_2 & 0 & \lambda_{e \tau } v_1\\
 \lambda_{\mu e} v_1 & \lambda_\mu v_2 & 0 \\
 0 & 0 & \lambda_\tau v_2 
\end{pmatrix}
~,\qquad
{M}_R =
\begin{pmatrix}
 M_{ee} & 0 & 0 \\
0 & 0 & M_{\mu\tau} \\
0 & M_{\mu\tau} & 0
\end{pmatrix}
~,
\label{eq:mdrdmi}
\end{equation}
while for the charged lepton mass matrix we have
\begin{equation}
 M_\ell =  \frac{1}{\sqrt{2}}
\begin{pmatrix}
 y_e v_2 & y_{e \mu } v_1& 0\\
 0 & y_\mu v_2 & 0 \\
 y_{\tau e} v_1 & 0 & y_\tau v_2 
\end{pmatrix}
~.
\label{eq:mldmi}
\end{equation}
Again there are off-diagonal components in $M_\ell$, whose implications
for the lepton-flavor violating processes will be discussed in
Sec.~\ref{sec:lfv}.

As before, there is an upper limit on the U(1)$_{L_\mu-L_\tau}$-symmetry
breaking scale since $v_1$ should be below the electroweak scale, and
thus a light gauge boson is again predicted in this case.

\section{Lepton flavor violating decay of charged leptons}
\label{sec:lfv}

As we see in Eqs.~\eqref{eq:mldpl} and \eqref{eq:mldmi}, in the doublet
cases the charged lepton mass matrix is not diagonal. It is diagonalized
by using unitary matrices $U_L$ and $U_R$ as 
\begin{equation}
 M_\ell = U_L^* 
\begin{pmatrix}
 m_e & 0 & 0\\
 0 & m_\mu & 0 \\
 0 & 0 & m_\tau 
\end{pmatrix}
U_R^T ~,
\label{eq:mldiag}
\end{equation}
where the gauge eigenstates $\ell_{L, R}$ are related to the mass
eigenstates $\ell^\prime_{L, R}$ as $\ell_{L, R} = U_{L, R}
\ell^\prime_{L, R}$. In the mass eigenbasis, the interactions of the
U(1)$_{L_\mu-L_\tau}$ gauge boson with the charged leptons are given by
\begin{equation}
 {\cal L}_{Z^\prime}
= g_{Z^\prime} \overline{\ell^\prime} \gamma^\mu 
\left[
U_L^\dagger Q_{\mu - \tau} U_L P_L
+ U_R^\dagger Q_{\mu - \tau} U_R P_R
\right] \ell^\prime Z_\mu^\prime ~,
\label{eq:lagzpr}
\end{equation}
where $P_{L/R} = (1\mp \gamma_5)/2$, $Z^\prime_\mu$ denotes the
U(1)$_{L_\mu-L_\tau}$ gauge field, and 
\begin{equation}
 \ell^\prime = 
\begin{pmatrix}
 e^\prime \\ \mu^\prime \\ \tau^\prime
\end{pmatrix}
~, \qquad
Q_{\mu - \tau}
=
\begin{pmatrix}
 0 & 0 & 0\\
 0 & 1 & 0 \\
 0 & 0 & -1
\end{pmatrix}
~.
\end{equation}
We see that the interaction in Eq.~\eqref{eq:lagzpr} in general induces
flavor mixings in the charged lepton sector. The
lepton-flavor-violating processes are severely constrained by
experiments, which thus give stringent limits on such mixing.

As discussed in the previous section,
the U(1)$_{L_\mu-L_\tau}$-symmetry breaking scale in the doublet cases
should be below the electroweak scale. Moreover, to evade the experimental limits
such as the neutrino trident bound \cite{Geiregat:1990gz,
Mishra:1991bv, Altmannshofer:2014cfa, Altmannshofer:2014pba}, we need
$g_{Z^\prime} \lesssim 10^{-2}$ for $v_1 \lesssim 100$~GeV. As a
consequence, $m_{Z^\prime} \lesssim m_\tau$ is generically expected. 
In this case, the $\tau \to e Z^\prime$ decay occurs if the (1,3) component of the $Z'$-coupling in Eq.~\eqref{eq:lagzpr} is nonzero.
The partial decay width of this channel is computed as 
\begin{equation}
 \Gamma (\tau \to e Z^\prime)
= \frac{g_{Z^\prime}^2 m_\tau}{32\pi} 
\bigl[\bigl|\bigl(U_L^\dagger Q_{\mu-\tau} U_L\bigr)_{13}
\bigr|^2 + \bigl|\bigl(U_R^\dagger Q_{\mu-\tau} U_R\bigr)_{13}
\bigr|^2\bigr] \biggl(2+ \frac{m_\tau^2}{m_{Z^\prime}^2}\biggr)
\biggl(1-\frac{m_{Z^\prime}^2}{m_\tau^2}\biggr)^2 ~,
\label{eq:gamtauez}
\end{equation}
where we have neglected the electron mass. Notice that when
$m_{Z^\prime} \ll m_\tau$, the decay width is enhanced by a factor
$m_\tau^2/m_{Z^\prime}^2$; this enhancement originates from the
longitudinal component of $Z^\prime$ in the final state.
For the $\mu \to eZ^\prime$
channel, the corresponding expression can be obtained by replacing $(U_{L/R}^\dagger Q_{\mu-\tau}U_{L/R})_{13}$
with $(U_{L/R}^\dagger Q_{\mu-\tau}U_{L/R})_{12}$ and $\tau$ with $\mu$ in Eq.~\eqref{eq:gamtauez}.

To see how strong the limits from the lepton-flavor-violating processes are,
let us consider the case (ii) with $M_\ell$ in
Eq.~\eqref{eq:mldpl}, and focus on the $\tau \to e Z^\prime$ channel as
an example. To simplify the discussion, we set $y_{\mu e} = 0$ and $y_\mu v_2/\sqrt{2}=m_\mu$, and
examine the effect of $y_{e\tau}$. We can always take $y_{e} v_2$,
$y_{\mu} v_2$, and $y_{\tau} v_2$ to be real and positive without loss
of generality. In this basis, $y_{e\tau} v_1$ is in general complex. The
unitary matrices $U_L$ and $U_R$ in Eq.~\eqref{eq:mldiag} are then
parametrized as follows:
\begin{equation}
 U_{L, R} = 
\begin{pmatrix}
 \cos \theta_{L,R} & 0 & e^{-i\phi} \sin \theta_{L,R} \\
 0 & 1 & 0 \\
 - e^{i\phi} \sin \theta_{L,R} & 0 & \cos \theta_{L,R}
\end{pmatrix}
~,
\label{eq:13mixing}
\end{equation}
where $\phi = \text{arg}(y_{e\tau} v_1)$ and 
\begin{equation}
 \frac{\tan \theta_R}{\tan \theta_L} = \frac{m_e}{m_\tau} ~.
\end{equation}
The mixing angle is related to the off-diagonal component through the
following equation:
\begin{equation}
 \left|y_{e\tau} v_1\right|
= \frac{(m_\tau^2 -m_e^2) \sin 2\theta_L}{\sqrt{
(m_\tau^2 + m_e^2) + (m_\tau^2 - m_e^2) \cos 2 \theta_L
}} ~.
\end{equation}
Using this mixing angle, the decay width of the $\tau \to eZ^\prime$
channel in Eq.~\eqref{eq:gamtauez} is expressed as 
\begin{equation}
 \Gamma (\tau \to eZ^\prime) = 
\frac{g_{Z^\prime}^2 m_\tau}{128\pi} 
\sin^2 2 \theta_L 
\biggl(2+ \frac{m_\tau^2}{m_{Z^\prime}^2}\biggr)
\biggl(1-\frac{m_{Z^\prime}^2}{m_\tau^2}\biggr)^2 ~. 
\label{eq:gamtauezsimp}
\end{equation}

On the other hand, there is an experimental upper limit on the two-body
decay of $\tau$ into an electron and a missing particle imposed by the
ARGUS Collaboration \cite{Albrecht:1995ht}. If the mass of the missing
particle $X$ is smaller than about $500$~MeV, the limit is
$\text{BR}(\tau \to eX)/\text{BR} (\tau \to e \nu \bar{\nu}) \lesssim
0.015$, with $\text{BR} (\tau \to e \nu \bar{\nu}) = 0.1782(4)$
\cite{Tanabashi:2018oca}. For a larger mass of $X$, the limit gets
weaker---the weakest bound is $\text{BR}(\tau \to eX)/\text{BR} (\tau \to
e \nu \bar{\nu}) \lesssim 0.035$ for an $X$ mass of $\sim 1$~GeV---and
then more stringent limits are set for masses larger than 1 GeV up to
$1.6$~GeV. For $m_{Z^\prime} < 2 m_\mu$, $Z^\prime$ dominantly decays into
neutrinos and thus it is invisible in experiments. Therefore, we can
directly apply the ARGUS limit, $\text{BR}(\tau \to eX) \lesssim 2.7
\times 10^{-3}$, in this case. By using Eq.~\eqref{eq:gamtauezsimp} as
well as the lifetime of $\tau$, $(290.3 \pm 0.5) \times 10^{-15}$~s
\cite{Tanabashi:2018oca}, we obtain a
limit on the mixing angle $\theta_L$ from the ARGUS limit as 
\begin{equation}
 |\sin 2 \theta_L| < 
\begin{cases}
 7 \times 10^{-5}
 & \text{for}~m_{Z^\prime} = 100~\text{MeV}~\text{and} ~g_{Z^\prime} =
 10^{-3} \\
 1 \times 10^{-5}
 &  \text{for}~m_{Z^\prime} = 10~\text{MeV}~\text{and} ~g_{Z^\prime} =
 5 \times 10^{-4}
\end{cases}
~.
\end{equation}
This shows that the mixing angle should be extremely close to either $0$ or
$\pi/2$. Note that this limit remains quite strong even if we take
$g_{Z^\prime}$ to be very small. In this case, $m_{Z^\prime}$ also
gets small, and $\Gamma (\tau \to eZ^\prime)$ goes as $\propto
g_{Z^\prime}^2/m_{Z^\prime}^2 \sim 1/v_1^2$, remaining constant.  
The $\tau$-$e$ mixing for the case (iii), induced by the off-diagonal component in
Eq.~\eqref{eq:mldmi}, is also constrained by the ARGUS limit in a similar
manner. Even if the two-body decay processes are kinematically forbidden,
the three-body lepton-flavor changing decay processes can still occur,
such as $\tau^- \to e^- \mu^+ \mu^-$. The
present limit on this decay mode is $\text{BR}(\tau^- \to e^- \mu^+
\mu^-) < 2.7 \times 10^{-8}$ \cite{Hayasaka:2010np}, which is found to
constrain the mixing angle at the ${\cal O}(10^{-(3-5)})$ level,
depending on the mass of $Z^\prime$. 
This limit is also applicable for $2m_\mu<m_{Z^\prime}\lesssim m_\tau$, where the two-body decay process $\tau \to e Z^\prime$ is
allowed and accompanied by $Z^\prime \to \mu^+ \mu^-$,
and it again results in a very strong limit on the mixing
angle. The limit on the $\tau \to e \gamma$ channel, 
$\text{BR}(\tau \to e \gamma) < 3.3 \times 10^{-8}$ 
\cite{Aubert:2009ag}, also gives a severe constraint.  
We thus conclude that the
$\tau$-$e$ mixing should be strongly suppressed in the doublet
scenarios. 

For the $\mu$-$e$ mixing induced by the (1,2) component of the $Z'$-coupling in Eq.~\eqref{eq:lagzpr}, we
may use the limit on the $\mu \to e X$ decay if $\mu \to eZ^\prime$ is
kinematically allowed. Currently, the most stringent limit on this decay
channel is $\text{BR}(\mu \to eX)/\text{BR} (\mu \to e \nu
\bar{\nu}) < 2.6 \times 10^{-6}$ for a massless $Z^\prime$
\cite{Jodidio:1986mz}; a similarly strong limit is obtained for
$m_{Z^\prime} \lesssim 16$~MeV \cite{Jodidio:1986mz}. The TWIST
collaboration also gives an upper limit, $\text{BR}(\mu \to eX) \lesssim
10^{-5}$ for $m_{Z^\prime} = 13$--80~MeV \cite{Bayes:2014lxz}. For heavier $Z^\prime$, the limit gets weaker to be $\lesssim 10^{-4}$
\cite{Bryman:1986wn}. In addition to this direct two-body decay channel,
$Z^\prime$ can also give rise to $\mu \to e\gamma$ at loop level through
kinetic mixing of $Z^\prime$ with $\gamma$ induced by the $\mu$ and
$\tau$ loops. For this decay channel, an extremely strong limit is
obtained by the MEG Experiment: $\text{BR}(\mu \to e\gamma) < 4.2 \times
10^{-13}$ \cite{TheMEG:2016wtm}. In any cases, the $\mu$-$e$ mixing is
again severely restricted.

As a consequence, we are forced to make the charged lepton-flavor mixing
extremely small in the cases (ii) and (iii). For the $e$-$\tau$ mixing, this means $\theta_L = 0$ or $\pi/2$ in Eq.~\eqref{eq:13mixing}.
$\theta_L = 0$ merely indicates $M_\ell = \text{diag}(m_e,
m_\mu, m_\tau)$ as $U_{L,R} = \1$. For $\theta_L = \pi/2$, on the other
hand, we have
\begin{equation}
 U_{L, R} = 
\begin{pmatrix}
  0 & 0 & e^{-i\phi}  \\
 0 & 1 & 0 \\
 - e^{i\phi}  & 0 & 0
\end{pmatrix}
= 
\begin{pmatrix}
  0 & 0 & 1  \\
 0 & 1 & 0 \\
 1  & 0 & 0
\end{pmatrix}
\begin{pmatrix}
   - e^{i\phi} & 0 & 0  \\
 0 & 1 & 0 \\
0  & 0 & e^{-i\phi}
\end{pmatrix}
~,
\end{equation}
with which Eq.~\eqref{eq:mldiag} leads to
\begin{equation}
 M_\ell = 
\begin{pmatrix}
  0 & 0 & 1  \\
 0 & 1 & 0 \\
 1  & 0 & 0
\end{pmatrix}
\begin{pmatrix}
 m_e & 0 & 0\\
 0 & m_\mu & 0 \\
 0 & 0 & m_\tau 
\end{pmatrix}
\begin{pmatrix}
  0 & 0 & 1  \\
 0 & 1 & 0 \\
 1  & 0 & 0
\end{pmatrix}
~.
\end{equation}
This indicates that $U_{L}$ and $U_R$ in this case are equivalent to a
three-dimensional representation of an element in $S_3$. Similar
arguments can be applied to the other mixing cases. Hence, the general form
of $M_\ell$ that is free from the charged lepton flavor violation is
again given by Eq.~\eqref{eq:mlgeneric}.

We however note that if $g = g_{\tau  e \mu}$, $g_{\tau \mu e}$, $g_{\mu
e \tau}$, or $g_{\mu \tau e}$, the doublet models suffer from various
phenomenological constraints. As we see from Eq.~\eqref{eq:lagzpr}, these
cases are equivalent to either U(1)$_{L_e - L_\mu}$ or U(1)$_{L_e -
L_\tau}$ models in the mass eigenbasis. These gauge theories are severely restricted by various
experiments for $m_{Z^\prime} \lesssim 100$~GeV. In the doublet models
we have $m_{Z^\prime} = g_{Z^\prime} v_1$ with $v_1 \lesssim
100$~GeV, and it turns out that such a $Z^\prime$ is excluded in both
the U(1)$_{L_e - L_\mu}$ and U(1)$_{L_e - L_\tau}$ gauge models
\cite{Wise:2018rnb, Bauer:2018onh, Chun:2018ibr}. We therefore focus on
the $g = g_{e \mu \tau}$ and $g_{e \tau \mu}$ cases for the doublet models in what
follows.

\section{Neutrino mass and mixing structures}
\label{sec:neumass}

Next, we examine the neutrino mass and mixing structure in each
model. In particular, we see that there are two conditional equations
that should be satisfied by the low-energy neutrino parameters in each
model, which make four parameters among them dependent on the rest of parameters.

\subsection{Singlet}

As we mentioned above, we allow $M_\ell$ to have a generic form
\eqref{eq:mlgeneric}. Throughout this work, we assume that the
non-zero components in the Majorana mass matrix $M_R$ are much larger
than those in the neutrino Dirac matrix $M_D$ so that the mass matrix of
the active neutrinos is given by the seesaw formula
\begin{equation}
 M_\nu = - M_D^{} M_R^{-1} M_D^T ~,
\end{equation}
where $M_D$ and $M_R$ are given in Eq.~\eqref{eq:mdrs}. 
This mass matrix can be diagonalized using a unitary matrix $U_\nu$:
\begin{equation}
 U_\nu^T M_\nu^{} U_\nu^{} = \text{diag} (m_1, m_2, m_3) ~,
\label{eq:mndiagn}
\end{equation}
where $m_i$ $(i = 1,2,3)$ are the mass eigenvalues. This unitary matrix
is related to the Pontecorvo-Maki-Nakagawa-Sakata (PMNS) mixing matrix
\cite{Pontecorvo:1967fh, Pontecorvo:1957cp, Pontecorvo:1957qd,
Maki:1962mu} $U_{\text{PMNS}}$ by 
\begin{equation}
 U_{\text{PMNS}} = D_3^T (g) U_\nu^{} ~,
\label{eq:upmnsd3unu}
\end{equation}
where $D_3 (g)$ is given in Eq.~\eqref{eq:mlgeneric}.
We parametrize the PMNS matrix as
\begin{equation}
\begin{pmatrix}
 c_{12} c_{13} & s_{12} c_{13} & s_{13} e^{-i\delta} \\
 -s_{12} c_{23} -c_{12} s_{23} s_{13} e^{i\delta}
& c_{12} c_{23} -s_{12} s_{23} s_{13} e^{i\delta}
& s_{23} c_{13}\\
s_{12} s_{23} -c_{12} c_{23} s_{13} e^{i\delta}
& -c_{12} s_{23} -s_{12} c_{23} s_{13} e^{i\delta}
& c_{23} c_{13}
\end{pmatrix}
\begin{pmatrix}
 1 & & \\
 & e^{i\frac{\alpha_{2}}{2}} & \\
 & & e^{i\frac{\alpha_{3}}{2}}
\end{pmatrix}
~,
\end{equation}
where $c_{ij} \equiv \cos \theta_{ij}$ and $s_{ij} \equiv \sin
\theta_{ij}$ for $\theta_{ij} = [0, \pi/2]$, $\delta = [0, 2\pi]$,
and we have ordered $m_1<m_2$ without loss of generality. 
We follow the convention of the Particle Data Group~\cite{Tanabashi:2018oca}, 
where $m_2^2-m_1^2\ll |m_3^2-m_1^2|$ and $m_1<m_2<m_3$ for the normal ordering (NO) or
$m_3<m_1<m_2$ for the inverted ordering (IO).

As shown in Ref.~\cite{Asai:2017ryy}, $m_i$ ($i = 1,2,3$) should be
non-zero in order for $M_\nu$ not to be block-diagonal. Then, we can
invert Eq.~\eqref{eq:mndiagn} to obtain
\begin{equation}
 M_\nu^{-1} = U_\nu^{} \text{diag}(m_1^{-1}, m_2^{-1}, m_3^{-1}) U_\nu^T 
= - (M_D^{-1})^T M_R M_D^{-1} ~.
\end{equation}
Now in the singlet case, $M_D^{-1}$ is diagonal, and the $(\mu, \mu)$ and $(\tau, \tau)$ components of $M_R$ are zero (see
Eq.~\eqref{eq:mdrs}). It then follows from the above equation that the $(\mu, \mu)$ and $(\tau, \tau)$ components in
$M_\nu^{-1}$ are also zero---this 
type of structure of the neutrino mass matrix is dubbed as the two-zero minor
\cite{Lavoura:2004tu, Lashin:2007dm}. In particular, this specific
structure is called ${\bf C}^R$ in Ref.~\cite{Araki:2012ip}, where the
$(\mu, \mu)$ and $(\tau, \tau)$ components of the inverse of the
neutrino mass matrix vanish.
By using
Eq.~\eqref{eq:upmnsd3unu}, we can express this condition in
terms of the following two equations:
\begin{align}
  \left[D_3(g)^{} U_{\text{PMNS}}^{} \text{diag}(m_1^{-1}, m_2^{-1}, m_3^{-1})
  U_{\text{PMNS}}^T  D_3^T(g)\right]_{\mu\mu}  &= 0 ~,
\nonumber \\[3pt]
 \left[D_3(g)^{} U_{\text{PMNS}}^{} \text{diag}(m_1^{-1}, m_2^{-1}, m_3^{-1})
  U_{\text{PMNS}}^T  D_3^T(g)\right]_{\tau\tau} &= 0~.
\label{eq:coneqs}
\end{align}
The left-hand side of these equations are complex, so four real degrees
of freedom are constrained by these conditions. The parameters included
in these equations are $m_i$ ($i =1,2,3$), $\theta_{12}$, $\theta_{23}$,
$\theta_{13}$, $\delta$, $\alpha_2$, $\alpha_3$; among these nine parameters,
four independent linear combinations of them are regarded as dependent
on the other five degrees of freedom. In the following analysis, we take 
the two squared mass differences and the three mixing angles as input
parameters, and derive the values of $\delta$, $\alpha_2$, $\alpha_3$,
and $\sum_{i} m_i$ from the five input parameters. Some analytical
expressions that are useful to determine these values are given in
Ref.~\cite{Asai:2017ryy}.

Notice that the conditional equations in Eq.~\eqref{eq:coneqs} do not
contain the scale of the U(1)$_{L_\mu-L_\tau}$ symmetry breaking
explicitly. In addition, it is shown in Ref.~\cite{Asai:2017ryy} that
the two-zero minor structure remains unchanged under the renormalization
group flow when the charged-lepton Dirac Yukawa matrix is
diagonal. Therefore, the conclusion we draw in this subsection holds even if
the U(1)$_{L_\mu-L_\tau}$ symmetry breaking scale is much higher than
the electroweak scale, which is possible in the singlet case.

There are six cases in the
singlet model and each of them corresponds to a different element
$g_{{\ell_1}{\ell_2}{\ell_3}}$ of the symmetry group $S_3$, and thus a different $M_\ell =
\text{diag}(m_{\ell_1}, m_{\ell_2}, m_{\ell_3})$. Now we note that the
conditional equations in Eq.~\eqref{eq:coneqs} are invariant under the
exchange of $\mu$ and $\tau$. This corresponds to a transformation
$D_3(g_{{\ell_1}{\ell_2}{\ell_3}}) \to D_3
(g_{{e}{\tau}{\mu}}) D_3(g_{{\ell_1}{\ell_2}{\ell_3}}) =
D_3(g_{{\ell_1}{\ell_3}{\ell_2}})$, and thus the predictions in the case
$g_{{\ell_1}{\ell_2}{\ell_3}}$ are the same as those in the case
$g_{{\ell_1}{\ell_3}{\ell_2}}$. In other words, in terms of the diagonal
components of $M_\ell$,
\begin{itemize}
 \item The cases with $M_\ell = \text{diag}(m_e, m_\mu, m_\tau)$ and
       $\text{diag}(m_e, m_\tau, m_\mu)$;
 \item The cases with $M_\ell = \text{diag}(m_\mu, m_e, m_\tau)$ and
       $\text{diag}(m_\mu, m_\tau, m_e)$;
 \item The cases with $M_\ell = \text{diag}(m_\tau, m_e, m_\mu)$ and
       $\text{diag}(m_\tau, m_\mu, m_e)$;
\end{itemize}
are equivalent, respectively. As noted above, the second (third) case
corresponds to the U(1)$_{L_e-L_\tau}$ (U(1)$_{L_e-L_\mu}$) theory in
the mass eigenbasis.

\subsection{Doublet with the U(1)$_{L_\mu-L_\tau}$ charge $+1$}

Next, we discuss the neutrino mass structure resulting from $M_D$ and
$M_R$ in Eq.~\eqref{eq:mdrdpl}. By using the seesaw formula, we obtain
\begin{equation}
 M_\nu = -
\begin{pmatrix}
 \frac{\lambda_e^2 v_2^2}{2 M_{ee}} & 0 & 
\frac{\lambda_{e\mu} \lambda_\tau v_1 v_2}{2 M_{\mu\tau}}
+ \frac{\lambda_e \lambda_{\tau e} v_1 v_2}{2 M_{ee}} \\[2pt]
 0 & 0 & \frac{\lambda_\mu \lambda_\tau v_2^2}{2 M_{\mu\tau}} \\[2pt]
\frac{\lambda_{e\mu} \lambda_\tau v_1 v_2}{2 M_{\mu\tau}}
+ \frac{\lambda_e \lambda_{\tau e} v_1 v_2}{2 M_{ee}} & 
\frac{\lambda_\mu \lambda_\tau v_2^2}{2 M_{\mu\tau}}  &
 \frac{\lambda_{\tau e}^2 v_1^2}{2 M_{ee}}
\end{pmatrix}
~.
\label{eq:mnudpl}
\end{equation}
This has a structure called the two-zero texture \cite{Berger:2000zj,
Frampton:2002yf, Xing:2002ta, Kageyama:2002zw, Xing:2002ap}, and
denoted by ${\bf B}_3^\nu$ in Ref.~\cite{Araki:2012ip}. This mass matrix
is diagonalized in a similar way to Eq.~\eqref{eq:mndiagn}:
\begin{equation}
 M_\nu = U^*_\nu \text{diag} (m_1, m_2, m_3) U_\nu^\dagger 
= D_3(g) U^*_{\text{PMNS}} 
\text{diag} (m_1, m_2, m_3)
U^\dagger_{\text{PMNS}} D^T_3 (g) ~,
\label{eq:mndiagn2}
\end{equation}
where we have used Eq.~\eqref{eq:upmnsd3unu}, and $g = g_{e \mu \tau}$
or $g_{e \tau \mu}$. The conditional equations
in this case are obtained from the $(e, \mu)$ and $(\mu, \mu)$
components in the above equation:
\begin{align}
 \left[D_3(g) U^*_{\text{PMNS}} 
\text{diag} (m_1, m_2, m_3)
U^\dagger_{\text{PMNS}} D^T_3 (g)\right]_{e\mu} &= 0 ~, 
\nonumber \\[3pt]
\left[D_3(g) U^*_{\text{PMNS}} 
\text{diag} (m_1, m_2, m_3)
U^\dagger_{\text{PMNS}} D^T_3 (g)\right]_{\mu \mu} &= 0 ~.
\label{eq:coneqdpl}
\end{align}
Again, we can determine the four
parameters $\delta$, $\alpha_2$, $\alpha_3$, and $\sum_{i} m_i$ 
as functions of the neutrino oscillation parameters from these equations.

\subsection{Doublet with the U(1)$_{L_\mu-L_\tau}$ charge $-1$}

As for $M_D$ and $M_R$ in Eq.~\eqref{eq:mdrdmi}, we have 
\begin{equation}
 M_\nu = -
\begin{pmatrix}
 \frac{\lambda_e^2 v_2^2}{2 M_{ee}} & 
\frac{\lambda_{e\tau} \lambda_\mu v_1 v_2}{2 M_{\mu\tau}}
+ \frac{\lambda_e \lambda_{\mu e} v_1 v_2}{2 M_{ee}}
& 
0 \\[2pt]
 \frac{\lambda_{e\tau} \lambda_\mu v_1 v_2}{2 M_{\mu\tau}}
+ \frac{\lambda_e \lambda_{\mu e} v_1 v_2}{2 M_{ee}}
 & 
 \frac{\lambda_{\mu e}^2 v_1^2}{2 M_{ee}}
& \frac{\lambda_\mu \lambda_\tau v_2^2}{2 M_{\mu\tau}} \\[2pt]
0 & 
\frac{\lambda_\mu \lambda_\tau v_2^2}{2 M_{\mu\tau}}  &
0
\end{pmatrix}
~.
\end{equation}
Again, this has a form of the two-zero texture, denoted by ${\bf B}_4^\nu$
in Ref.~\cite{Araki:2012ip}.
By using Eq.~\eqref{eq:mndiagn2} and taking the $(e, \tau)$ and
$(\tau, \tau)$ components, we obtain
\begin{align}
 \left[D_3(g) U^*_{\text{PMNS}} 
\text{diag} (m_1, m_2, m_3)
U^\dagger_{\text{PMNS}} D^T_3 (g)\right]_{e\tau} &= 0 ~, 
\nonumber \\[3pt]
\left[D_3(g) U^*_{\text{PMNS}} 
\text{diag} (m_1, m_2, m_3)
U^\dagger_{\text{PMNS}} D^T_3 (g)\right]_{\tau \tau} &= 0 ~,
\label{eq:coneqdmi}
\end{align}
with $g = g_{e \mu \tau}$ or $g_{e \tau \mu}$.
These are the conditional equations for the model (iii). 

Notice that the conditions in Eq.~\eqref{eq:coneqdmi} are converted into
those in Eq.~\eqref{eq:coneqdpl} via the interchange of $\mu$ and
$\tau$. As a result, the cases specified by $M_\ell
= \text{diag}(m_{e}, m_{\mu}, m_{\tau})$ and $\text{diag}(m_{e},
m_{\tau}, m_{\mu})$ in the model (ii) make the same predictions as those
in the cases with $M_\ell = \text{diag}(m_{e}, m_{\tau}, m_{\mu})$ and
$\text{diag}(m_{e}, m_{\mu}, m_{\tau})$ in the model (iii), respectively.

\subsection{Summary}

\begin{table}[t]
 \begin{center}
\caption{The neutrino mass structures in the minimal gauged
  U(1)$_{L_\mu-L_\tau}$ models.  } 
\label{tab:structure}
\vspace{5pt}
\begin{tabular}{ccclc}
\hline
\hline
 Model & SU(2)$_L$ & U(1)$_{L_\mu-L_\tau}$ & Structure & Condition  \\
\hline
(i) & Singlet & $+1$ & Two-zero minor ${\bf C}^R$ & Eq.~\eqref{eq:coneqs} \\
(ii) & Doublet & $+1$ & Two-zero texture ${\bf B}^\nu_3$ & Eq.~\eqref{eq:coneqdpl} \\
(iii) & Doublet & $-1$ & Two-zero texture ${\bf B}^\nu_4$ & Eq.~\eqref{eq:coneqdmi} \\
\hline
\hline
\end{tabular}
 \end{center}
\end{table}

All in all, the neutrino mass structures found in the three models are
summarized in Table~\ref{tab:structure}. Each model is specified with
the quantum numbers of the U(1)$_{L_\mu-L_\tau}$-breaking scalar
field. We use the notation adopted in Ref.~\cite{Araki:2012ip} to
identify the neutrino mass structure. The equation numbers of the
resultant conditional expressions are also shown, which we use to
predict the values of $\sum_{i} m_i$ and the CP phases in the
subsequent section. For the model (i), there are three independent cases
and the rest three are equivalent to the former; while for each of the
two cases in the model (ii), there exists a case in the model (iii) that has
the same predictions. We will focus on the model (ii) for the doublet
cases in the following analysis. As a result, we have five independent
(three for the singlet model and two for the doublet models) cases to be
investigated.

\section{Neutrino phenomenology}
\label{sec:limit}

\begin{table}[t]
 \begin{center}
\caption{Values for the neutrino oscillation parameters we use
  in this paper. We take them from the \texttt{NuFIT v4.0} result
  with the Super-Kamiokande atmospheric data
 \cite{nufit, Esteban:2018azc}.  }
\label{tab:input}
\vspace{5pt}
\begin{tabular}{l|cc|cc}
\hline
\hline
 & \multicolumn{2}{c|}{Normal Ordering} &
 \multicolumn{2}{c}{Inverted Ordering} \\ 
\cline{2-5}
Parameter  & Best fit $\pm 1\sigma$& 3$\sigma$ range \qquad& Best fit
 $\pm 1\sigma$& 3$\sigma$ range \qquad  \\
\hline
$\sin^2 \theta_{12}$ & $0.310^{+0.013}_{-0.012}$ & 0.275--0.350 &
 $0.310^{+0.013}_{-0.012}$ & 0.275--0.350 \\
$\sin^2 \theta_{23}$ & $0.582^{+0.015}_{-0.019}$ & 0.428--0.624 &
 $0.582^{+0.015}_{-0.018}$ & 0.433--0.623 \\
$\sin^2 \theta_{13}$ & $0.02240^{+0.00065}_{-0.00066}$ & 0.02044--0.02437 & $0.02263^{+0.00065}_{-0.00066}$
 & 0.02067--0.02461 \\
$\Delta m^2_{21}/10^{-5}~\text{eV}^2$ & $7.39^{+0.21}_{-0.20}$ &
	 6.79--8.01 & $7.39^{+0.21}_{-0.20}$ & 6.79--8.01  \\
$\Delta m^2_{3\ell}/10^{-3}~\text{eV}^2$ & $2.525^{+0.033}_{-0.031}$ &
	 $2.431$--$2.622$ 
	 &$-2.512^{+0.034}_{-0.031}$ & $-$(2.606--$2.413$) \\
$\delta ~[{}^\circ]$ & $217^{+40}_{-28}$ & 135--366 & $280^{+25}_{-28}$ & 196--351\\
\hline
\hline
\end{tabular}
 \end{center}
\end{table}

Now we evaluate the values of $\sum_{i} m_i$ predicted in the minimal
gauged U(1)$_{L_\alpha-L_\beta}$ models. To that end, we regard the PMNS
mixing angles and the squared mass differences as input
parameters. These values are taken from the recent global fit,
\texttt{NuFIT v4.0} \cite{nufit, Esteban:2018azc}, which we list in
Table~\ref{tab:input}. Here, we take $\ell = 1$ for NO and $\ell = 2$
for IO in $\Delta m^2_{3\ell}$ \cite{Esteban:2016qun}. We also show the
favored value of the Dirac CP phase $\delta$, which is to be compared
with the values predicted in each model.

Let us first analyze the singlet cases. There are three independent
cases: 
(a)
$M_\ell = \text{diag}(m_e, m_\mu, m_\tau)$ 
or $\text{diag}(m_e, m_\tau, m_\mu)$; 
(b) 
$M_\ell = \text{diag}(m_\mu, m_e,m_\tau)$ 
or $\text{diag}(m_\mu, m_\tau, m_e)$; 
(c) $M_\ell = \text{diag}(m_\tau, m_e, m_\mu)$ 
or $\text{diag}(m_\tau, m_\mu, m_e)$. 
We study each case assuming either NO or IO, and solve the
conditional equations in Eq.~\eqref{eq:coneqs} to obtain $\sum_{i} m_i$
and the CP phases, using the corresponding parameter set in
Table~\ref{tab:input}. We then find that only the case (a) with NO has a
reasonable solution---the others have no solution for $\delta$ or the
resultant mass ordering is inconsistent with the assumption. This
is consistent with the conclusion drawn in 
Ref.~\cite{Asai:2017ryy}.

\begin{figure}[t]
\centering
\includegraphics[clip, width = 0.45 \textwidth]{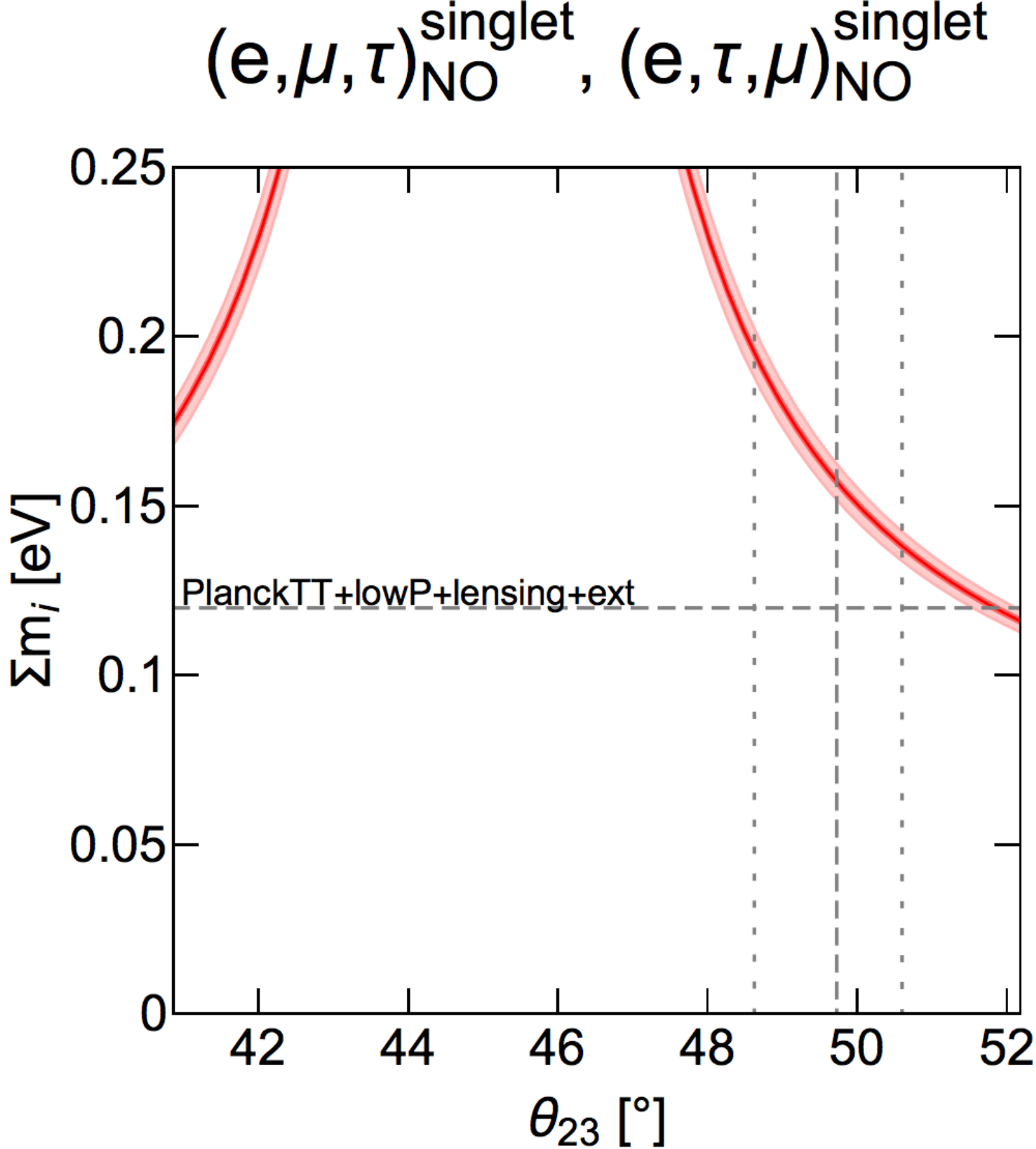}
\caption{The sum of the neutrino masses as a function of
 $\theta_{23}$ predicted in the singlet model with $M_\ell =
 \text{diag}(m_e, m_\mu, m_\tau)$ or $\text{diag}(m_e, m_\tau, m_\mu)$
 for NO. The vertical gray dashed line represents the best fit value of
 $\theta_{23}$, while the vertical gray dotted lines (the plot range) indicate
 the $1\sigma$ ($3\sigma$) range. 
 The dark (light) red band represents the uncertainty
coming from the $1\sigma$ ($3\sigma$) range of $\theta_{13}$.
We also show in the horizontal gray dashed line
 the limit imposed by the Planck experiment: $\sum_{i} m_i < 0.12$~eV
 (Planck TT+lowP+lensing+ext)~\cite{Aghanim:2018eyx}.}
\label{fig:sums}
\end{figure}

In Fig.~\ref{fig:sums}, we plot the sum of the neutrino masses as a
function of $\theta_{23}$ predicted in the case (a) with NO. The vertical
gray dashed line represents the best fit value of $\theta_{23}$, while
the vertical gray dotted lines (the plot range) indicate the $1\sigma$
($3\sigma$) range. 
The dark (light) red band represents the uncertainty
coming from the $1\sigma$ ($3\sigma$) range of $\theta_{13}$.
The effects of the other parameters' uncertainties are subdominant. 
We also show in the horizontal gray dashed line the limit
imposed by the Planck experiment: $\sum_{i} m_i < 0.12$~eV (Planck
TT+lowP+lensing+ext) \cite{Aghanim:2018eyx}. As we see, there is a
strong tension between the prediction and the Planck bound; the
predicted value barely avoids the limit only when we allow the
parameters to be varied in $3\sigma$. Hence, if the limit gets a
little bit more stringent in the future, then the singlet case will be
completely ruled out. We also note that such a large $\sum_{i} m_i$
implies a quasi-degenerate mass spectrum.

\begin{figure}
  \centering
  \subcaptionbox{\label{fig:sumdemtno} $M_\ell = \text{diag}(m_e, m_\mu,
 m_\tau)$ (NO)}{
  \includegraphics[width=0.35\columnwidth]{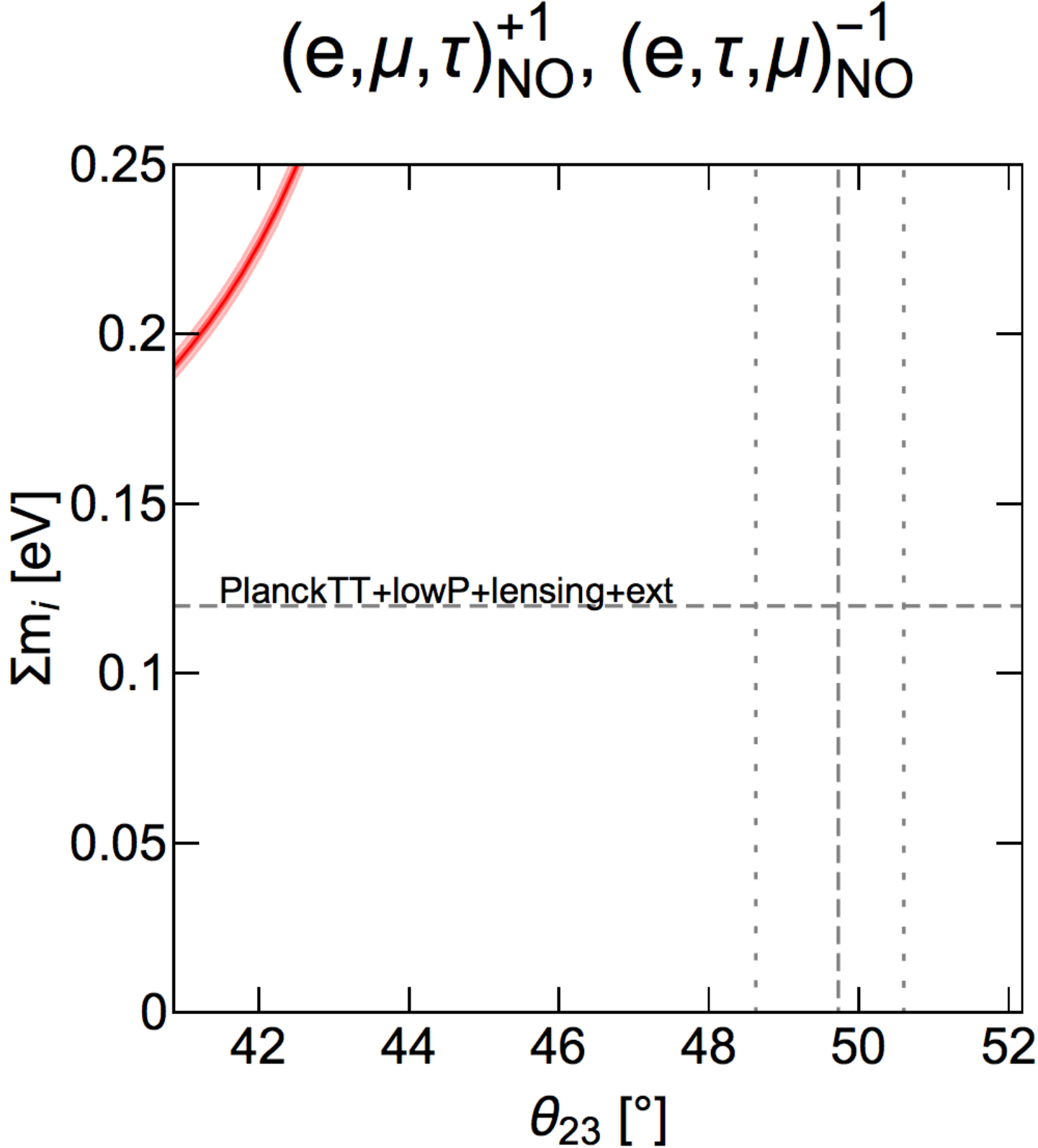}}
  \subcaptionbox{\label{fig:sumdemtio} $M_\ell = \text{diag}(m_e, m_\mu,
 m_\tau)$ (IO)}{
  \includegraphics[width=0.35\columnwidth]{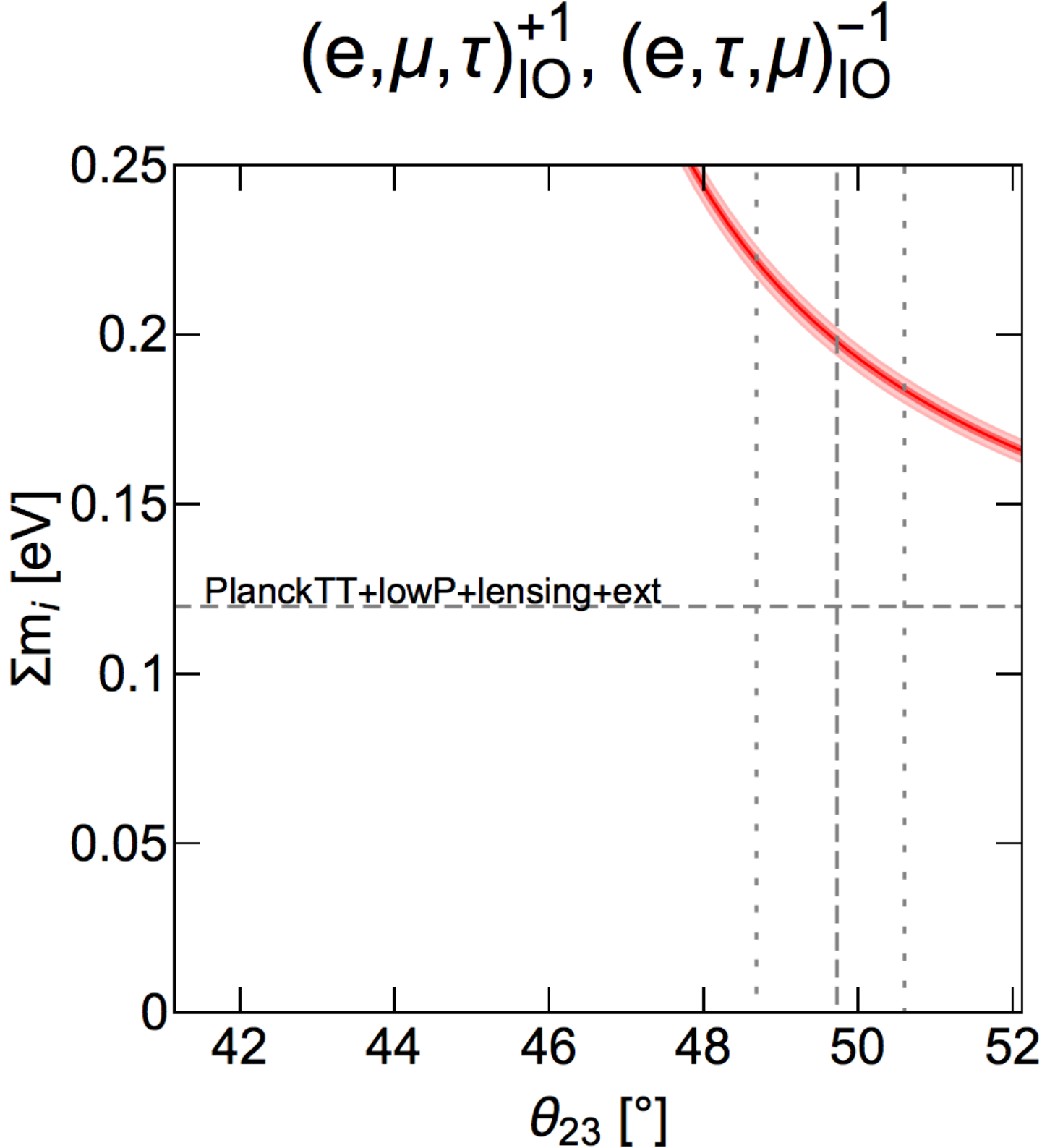}}\\
\vspace{5mm}
  \subcaptionbox{\label{fig:sumdetmno} $M_\ell = \text{diag}(m_e, m_\tau,
 m_\mu)$ (NO)}{
  \includegraphics[width=0.35\columnwidth]{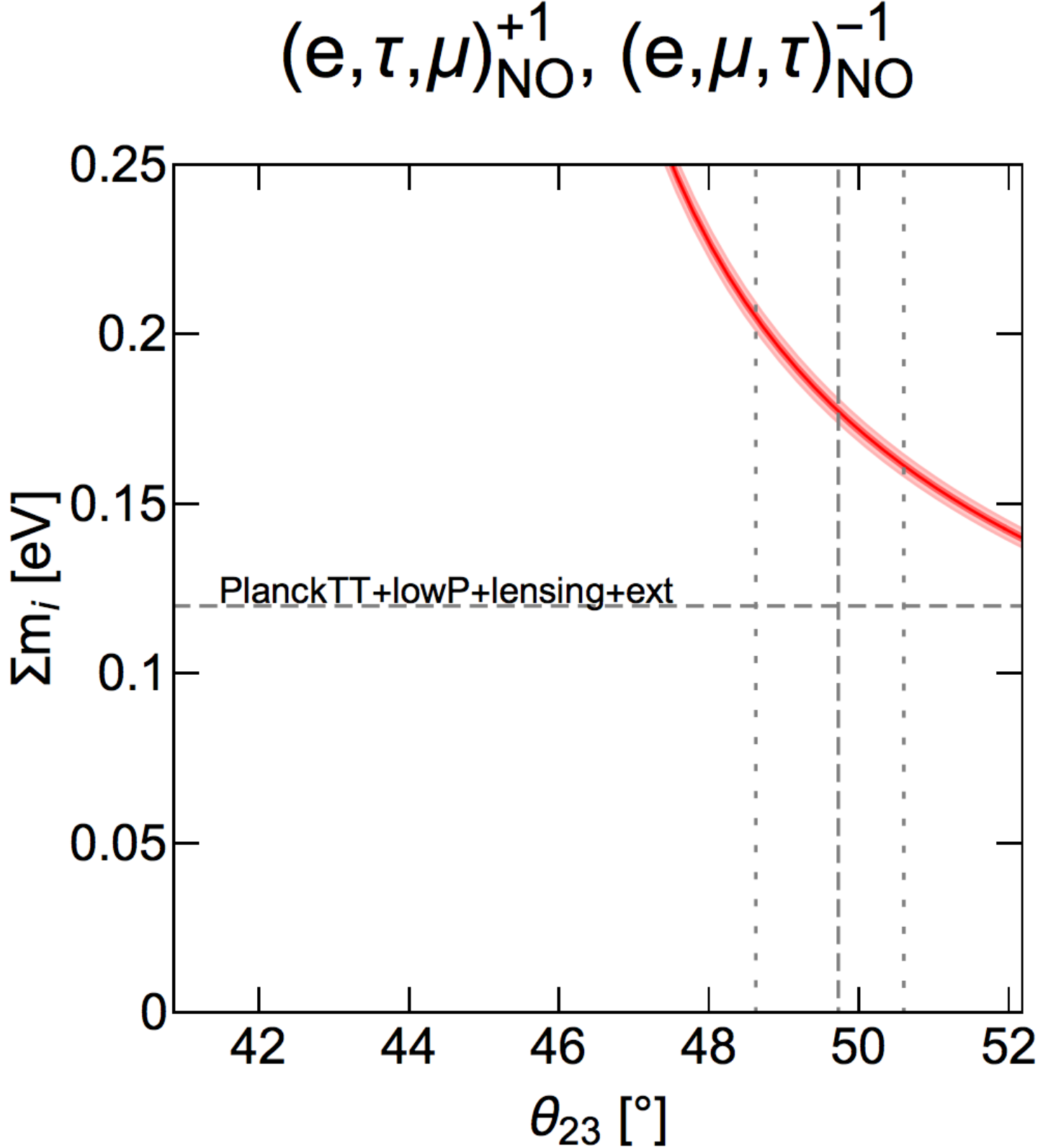}}
  \subcaptionbox{\label{fig:sumdetmio} $M_\ell = \text{diag}(m_e, m_\tau,
 m_\mu)$ (IO)}{
  \includegraphics[width=0.35\columnwidth]{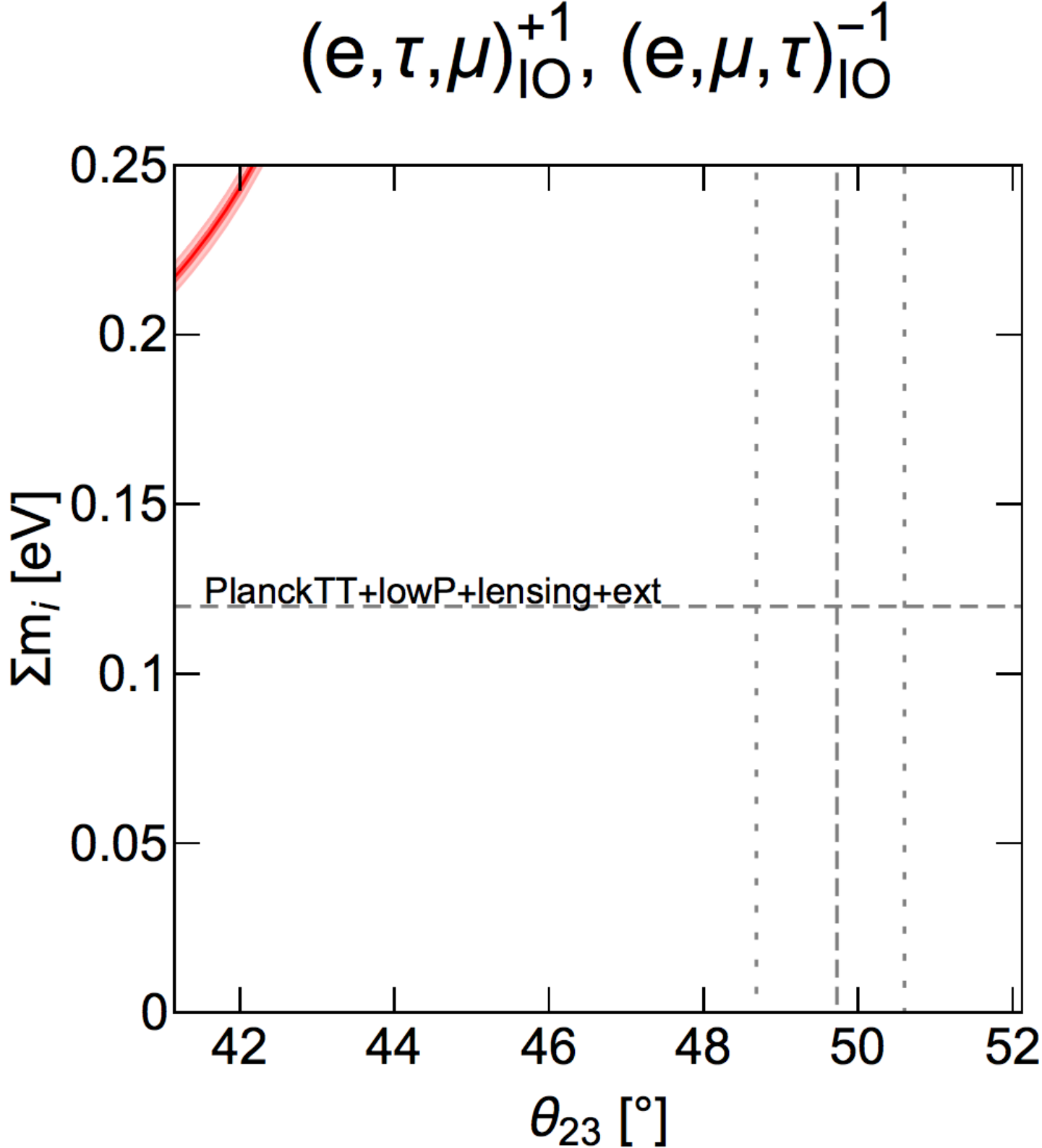}}
\caption{The sum of the neutrino masses as a function of
 $\theta_{23}$ predicted in the doublet models. The dark (light) red
 bands represent the $1\sigma$ ($3\sigma$) uncertainty coming from the
 $1\sigma$ ($3\sigma$) range of $\Delta m_{32}^2$. The vertical and
 horizontal lines are the same as in Fig.~\ref{fig:sums}.  } 
  \label{fig:sumd}
\end{figure}

For the doublet cases, we focus on the ones with the U(1)$_{L_\mu -
L_\tau}$ charge $+1$ as discussed above. We find that in the doublet
model a solution for the conditional expressions in
Eq.~\eqref{eq:coneqdpl} is obtained for all of the possible combinations between $g =
g_{e\mu \tau}, g_{e\tau \mu}$ and NO/IO.  
In Fig.~\ref{fig:sumd}, we show the predicted values of $\sum_{i} m_i$
as functions of $\theta_{23}$ for these four cases. 
The dark (light) red bands represent the uncertainty coming from the  
$1\sigma$ ($3\sigma$) range of $\Delta m_{32}^2$.
The effects of other parameters' uncertainties are subdominant. 
It turns out that all of these cases predict a too large $\sum_{i} m_i$ and
are excluded by the Planck limit. We can thus conclude that the minimal
gauged U(1)$_{L_\alpha - L_\beta}$ models with a doublet
U(1)$_{L_\alpha - L_\beta}$-breaking scalar have already been excluded.

By and large, there is basically only one possibility for the minimal
gauged U(1)$_{L_\alpha - L_\beta}$ models which are consistent with the
existing limits: the U(1)$_{L_\mu - L_\tau}$ model with a singlet
U(1)$_{L_\alpha - L_\beta}$-breaking scalar field, though this model is
also driven into a corner. We now study other 
predictions of this model and discuss the prospects of testing it in
future experiments.

\begin{figure}[t]
  \centering
  \subcaptionbox{\label{fig:deltas} Dirac CP phase}{
  \includegraphics[width=0.45\columnwidth]{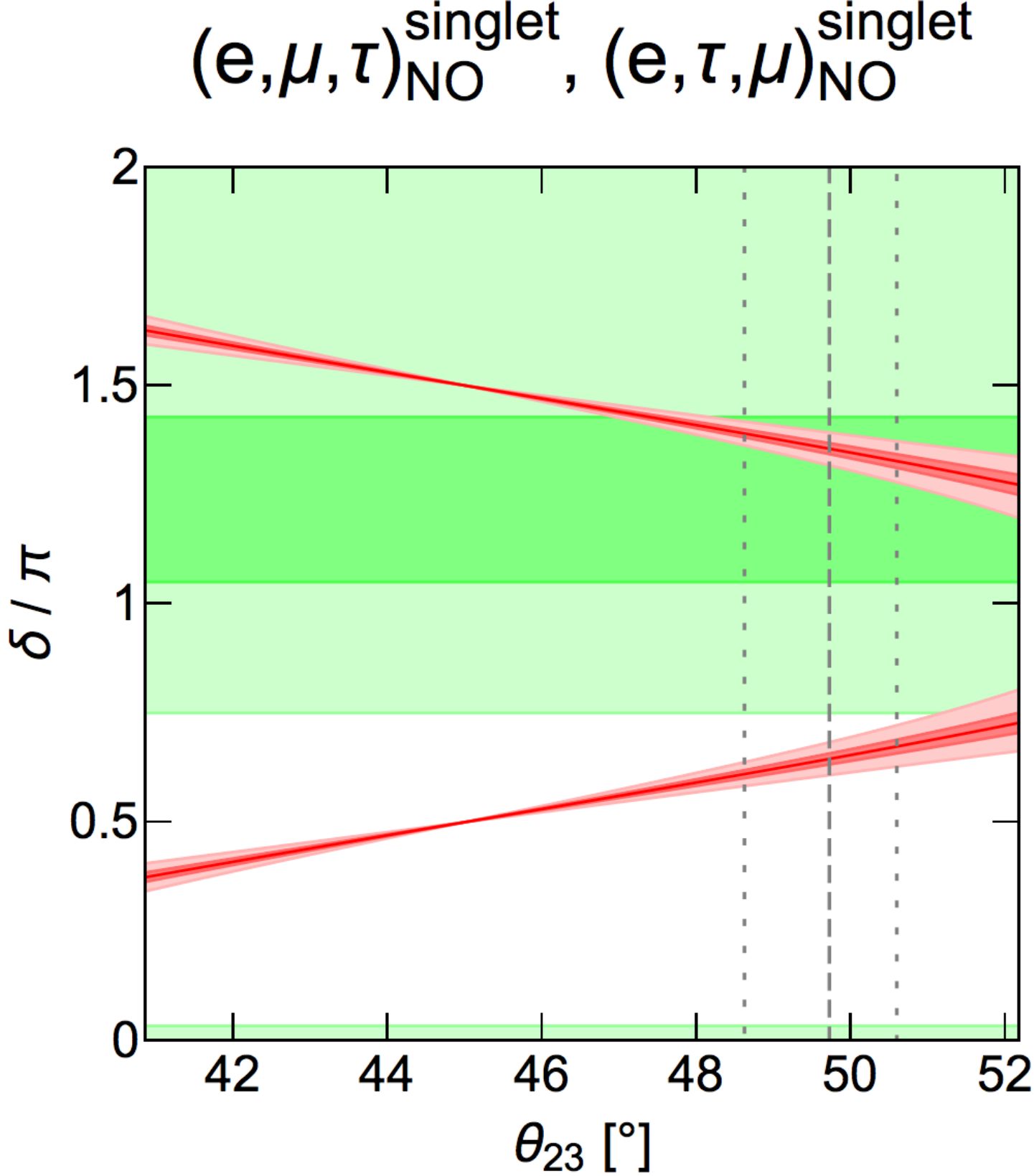}}
  \subcaptionbox{\label{fig:mbbs} Effective Majorana neutrino mass
 }{
  \includegraphics[width=0.45\columnwidth]{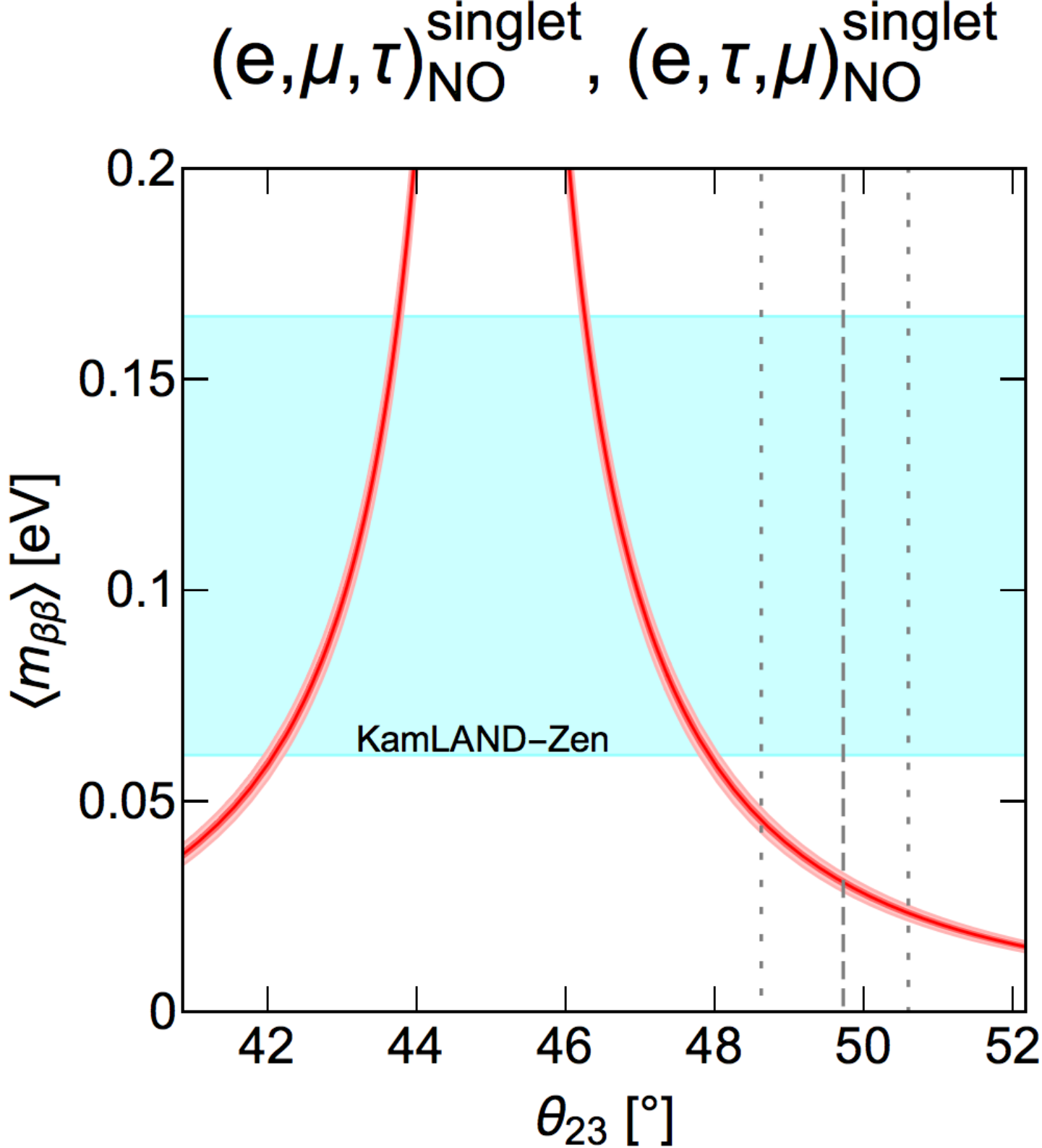}}
\caption{The predictions for (a) the Dirac CP phase $\delta$ 
and (b)
the effective
 Majorana neutrino mass $\langle m_{\beta\beta} \rangle$
 in the singlet case.
 The red lines show the predictions as functions of $\theta_{23}$, 
 and the dark (light) red bands show the uncertainty
 coming from the 1$\sigma$ ($3\sigma$) errors in the other parameters.
 The vertical gray dashed lines represent the best fit value of
 $\theta_{23}$, while the vertical gray dotted lines (the plot range) indicate
 the $1\sigma$ ($3\sigma$) range.
 In (a), we also
 show the 1$\sigma$ (3$\sigma$) favored region of $\delta$ in the dark
 (light) horizontal green bands.
 In (b), the light blue band represents the limit from KamLAND-Zen,
 $\langle m_{\beta\beta} \rangle <0.061$--0.165~eV~\cite{KamLAND-Zen:2016pfg}, where the band indicates uncertainty
 from the nuclear matrix element.
} 
  \label{fig:sing}
\end{figure}

First, in Fig.~\ref{fig:deltas}, we plot the Dirac CP phase $\delta$
versus $\theta_{23}$ in the red lines, with
the dark (light) red bands showing the uncertainty
 coming from the 1$\sigma$ ($3\sigma$) errors in $\theta_{12}$.
 The uncertainties from the other parameters are negligible.
We also
show the 1$\sigma$ (3$\sigma$) favored region of $\delta$ in the dark
(light) horizontal green bands. This figure shows that the predicted
value of $\delta$ falls right in the middle of the experimentally
favored range for $\theta_{23} \simeq 52^\circ$, around which
$\sum_{i} m_i \simeq 0.12$~eV as seen in Fig.~\ref{fig:sums}.

As suggested in the previous studies \cite{Crivellin:2015lwa,
Asai:2017ryy}, neutrinoless double-beta decay offers a promising way of
probing the singlet case. 
The rate of neutrinoless double-beta decay is proportional to the square
of the effective Majorana neutrino mass $\langle m_{\beta\beta}
\rangle$, which is defined by
\begin{align}
 \langle m_{\beta\beta} \rangle \equiv \biggl|
\sum_{i} \left(U_{\text{PMNS}}\right)_{ei}^2 m_i
\biggr|
=\left|
c_{12}^2 c_{13}^2 m_1 +
s_{12}^2 c_{13}^2 e^{i \alpha_2} m_2 +
s_{13}^2 e^{i(\alpha_3 -2\delta)} m_3 
\right| ~.
\end{align}
As all of the mass eigenvalues and both the Dirac and Majorana CP phases
are determined in the minimal gauged U(1)$_{L_\mu- L_\tau}$ models, the
value of the effective mass $\langle m_{\beta\beta} \rangle$ is also
determined unambiguously in terms of the oscillation parameters. We show the predicted value of $\langle
m_{\beta\beta} \rangle$ in Fig.~\ref{fig:mbbs} as a function of
$\theta_{23}$, where the dark (light) red band shows the uncertainty
coming from the 1$\sigma$ ($3\sigma$) errors in the parameters other than
$\theta_{23}$. We also show in the light blue band the current bound on
$\langle m_{\beta\beta} \rangle$ given by the KamLAND-Zen experiment,
$\langle m_{\beta\beta} \rangle <0.061$--0.165~eV
\cite{KamLAND-Zen:2016pfg}, where the uncertainty stems from the
estimation of the nuclear matrix element for $^{136}$Xe. We see that
$\langle m_{\beta\beta} \rangle$ is predicted to be $\simeq 0.016$~eV for
$\theta_{23} \simeq 52^\circ$, which is well below the present
KamLAND-Zen limit. Future experiments are expected to have sensitivities
as low as ${\cal O} (0.01)$~eV \cite{Agostini:2017jim}, and thus are
quite promising for testing this scenario.

In summary, the singlet case predicts 
\begin{itemize}
 \item Quasi-degenerate NO mass spectrum.
 \item $\sum_{i} m_i \gtrsim 0.12$~eV.
 \item $\theta_{23} \simeq 52^\circ$. 
 \item $\langle m_{\beta\beta} \rangle \gtrsim 0.016$~eV.
\end{itemize}
The measurements of these observables in future neutrino experiments
can verify or completely exclude the singlet scenario.

\section{Conclusion and discussion}
\label{sec:conclusion}

In this work, we have studied the neutrino mass structures of the
minimal gauged U(1)$_{L_\alpha-L_\beta}$ models in a systematic and
comprehensive manner. The neutrino mass matrices of these models have a
form of either two-zero minor or two-zero texture. Such a characteristic
structure requires the low-energy neutrino  parameters to obey two
conditional equations, which make four of them dependent on the rest of
the parameters. In particular, the sum of the neutrino masses is
predicted as a function of the neutrino oscillation parameters that are
measured with good accuracy in neutrino experiments. We then find that
most of the possible cases in the minimal gauged U(1)$_{L_\alpha-L_\beta}$
models are incompatible with the measured neutrino parameters or
excluded by the limit on $\sum_{i} m_i$ imposed by the Planck 2018
data. There remains only one possibility---the minimal gauged
U(1)$_{L_\mu-L_\tau}$ model with a singlet
U(1)$_{L_\mu-L_\tau}$-breaking field---though this is also forced into a
corner mainly due to the Planck 2018 limit on $\sum_{i} m_i$. Future measurements of
$\sum_{i} m_i$ and $\theta_{23}$, as well as the neutrino-less double-beta decay
experiments, can verify or exclude this model. 

It is pointed out in Ref.~\cite{Asai:2017ryy} that there is a 
non-trivial prediction for leptogenesis in the singlet model;
the asymmetry parameter for leptogenesis is unambiguously 
determined given a set of the neutrino Dirac Yukawa couplings.
In particular, since the positive (negative) sign of the asymmetry
parameter leads to the negative (positive) sign of baryon 
asymmetry of the Universe, the parameter space with a wrong-sign 
asymmetry parameter is then disfavored.
We performed the same analysis as in Ref.~\cite{Asai:2017ryy} 
with the up-to-date input parameters used in this paper and
found that in a wide range of parameter space, the asymmetry
parameter has the desired sign (negative), which makes leptogenesis
quite promising. This motivates a more detailed analysis on 
leptogenesis in the singlet scenario, which we defer to another 
opportunity.

Although we have focused on the minimal gauged U(1)$_{L_\alpha-L_\beta}$ models in
our work, a similar discussion can give interesting consequences for
other models. For example, the model based on the SU(2)$_{\mu\tau}$
gauge symmetry discussed in Ref.~\cite{Chiang:2017vcl} predicts the same
neutrino mass structure as in the singlet cases considered in this work, 
so the discussions given in this paper are also
applicable to this model. The same is the case with the model discussed in Refs.~\cite{Bian:2017rpg,Bian:2017xzg}. In Ref.~\cite{Dev:2017fdz}, an inverse seesaw
model with the gauged U(1)$_{L_\mu-L_\tau}$ symmetry is discussed, where
the neutrino mass matrix has a form of the two-zero texture. In this
case, the sum of the neutrino masses is predicted to be $\sum_{i}m_i
\gtrsim 0.15$~eV, and thus is in conflict with the Planck 2018
bound. The same two-zero mass structure is predicted in the models given
in Refs.~\cite{Baek:2015mna, Lee:2017ekw}, and thus these models also
suffer from the neutrino mass bound.


\section*{Acknowledgments}

This work is supported in part by the Grant-in-Aid for
Scientific Research A (No.16H02189 [KH]),
Young Scientists B (No.17K14270 [NN], No.16K17697 [KT]), Innovative Areas
(No.26104001 [KH], No.26104009 [KH], No.18H05542 [NN], No.18H05543 [KT]).


{\small 
\bibliographystyle{JHEP}
\bibliography{ref}
}

\end{document}